\documentclass[aps,prb,reprint,showpacs,superscriptaddress,amsmath,amssymb]{revtex4-1}% APS journal style [reprint->twocolumns]

\usepackage[utf8]{inputenc} % Required for non-English characters (input)
\usepackage[T1]{fontenc} % Required for non-English characters (output)
\usepackage[dvips]{epsfig}
\usepackage[english]{babel}
\usepackage[autostyle=true]{csquotes} % Generate language-dependent quotes in the bibliography
\usepackage[normalem]{ulem}
\usepackage{graphicx}% Include figure files
\usepackage{xcolor} %support colors
\usepackage{amsmath} % math formulas
\usepackage{amssymb} % math symbols
\usepackage{amsfonts} % math basic fonts
\usepackage{mathrsfs} % math RSFS fonts (e.g. \mathscr)
\usepackage{dsfont} % math double strike (e.g. integer set, unity matrix with \mathbb)
\usepackage{upgreek} % Upgreek symbols (e.g. \Upbeta)
\usepackage{bm} % bold math
\usepackage{placeins} % Control floats (e.g. \FloatBarrier)
\usepackage{comment}
\usepackage{float} % to use ['H'] in \includegraphics

\newcommand*{\opte}{{\em optical exciton}\,}
\newcommand*{\optes}{{\em optical excitons}\,}
\newcommand*{\unde}{{\em elemental exciton}\,}
\newcommand*{\undes}{{\em elemental excitons}\,}
\newcommand*{\OBSE}{{\em optical--}BSE\,}
\newcommand*{\EBSE}{{\em elemental--}BSE\,}

\newcommand*{\h}[1]{\hat{#1}}
\newcommand*{\di}{\mathrm{d}}

\newcommand*{\ttw}[2]{\(t_{#1},t_{#2}\)}
\newcommand*{\tth}[3]{\(t_{#1},t_{#2};t_{#3}\)}
\newcommand*{\tfo}[4]{\(t_{#1},t_{#2};t_{#3},t_{#4}\)}
\newcommand*{\mat}[1]{{\uu{#1}}}
\newcommand*{\tens}[1]{{\uu{\uu{#1}}}}
\newcommand*{\mind}[2]{_{\substack{#1\\#2}}}

\renewcommand{\[}{\left[}
\renewcommand{\]}{\right]}
\renewcommand{\(}{\left(}
\renewcommand{\)}{\right)}
\def\nl         {\right.\\ \left.}

\newcommand{\EQ}[1]{\begin{equation}#1\end{equation}}

\newcommand{\eq}[1]{\begin{align}#1\end{align}}
\newcommand{\ml}[1]{\begin{multline}#1\end{multline}}
\newcommand{\eqg}[1]{\begin{gather}#1\end{gather}}
\newcommand{\seq}[1]{\begin{subequations}#1\end{subequations}}
\newcommand{\stkout}[1]{\ifmmode\text{\sout{\ensuremath{#1}}}\else\sout{#1}\fi}
\newcommand{\braket}[2]{\left\langle #1 \right|\left. #2 \right\rangle}
\newcommand{\bra}[1]{\mbox{$\langle #1 |$}}
\newcommand{\ket}[1]{\mbox{$| #1 \rangle$}}
\newcommand{\average}[1]{\left\langle #1 \right\rangle}
\newcommand{\pr}[1]{\left( #1 \right)}

\newcommand{\lab}[1]{\label{#1}}

\newcommand{\mc}[1]{\mathcal{#1}}
\newcommand{\uu}[1]{\underline{#1}}
\newcommand{\oo}[1]{\overline{#1}}
\newcommand{\e}[1]{Eq.~\eqref{#1}}
\newcommand{\elab}[2]{Eq.~(\ref{#1}#2)}
\newcommand{\fig}[1]{Fig.~\ref{#1}}
\newcommand{\figlab}[2]{Fig.~\ref{#1}#2}

\newcommand{\app}[1]{Appendix\,\ref{#1}}

\newcommand{\dint}[1] {{\rm d}^3 #1}

\newcommand{\mr}[1]{{\mathrm #1}}

\def\im  {\mr{i}}            % Imaginary unit
\def\p   {\prime}
\def\gbel {\beta^{el}}
\def\gaopt {\alpha^{opt}}
\def\mos {MoS$_2$}
\def\mose {MoSe$_2$}

\def\ga         {\alpha}
\def\gb         {\beta}

\def\gC         {\Gamma}
\def\gd         {\delta}

\def\gee        {\epsilon}
\def\gl         {\lambda}

\def\go         {\omega}

\def\gr         {\rho}

\def\gS         {\Sigma}

\def\rr		{{\mathbf r}}

\def\kk		{{\mathbf k}}
\def\qq		{{\mathbf q}}

\def\CG {{\mc G}}
\def\ai	        {{\em ab initio}}

\newcommand{\cnrism} {Istituto di Struttura della Materia and Division of Ultrafast Processes in Materials (FLASHit) of the National Research Council, via Salaria Km 29.3, I-00016 Monterotondo Stazione, Italy}
\newcommand{\cnrnano} {CNR-NANO, Via Campi 213a, 41125 Modena, Italy}

\begin{document}

\title[Theory of exciton--phonon interaction]
  {Exciton--phonon interaction calls for a revision of the ``exciton'' concept}

\author{Fulvio Paleari}
\affiliation{\cnrism}
\affiliation{\cnrnano}
\email{fulvio.paleari@nano.cnr.it}

\author{Andrea Marini}
\affiliation{\cnrism}

\date{\today}

\begin{abstract}
The concept of \opte  --- a photo--excited bound electron--hole pair within a crystal --- is routinely used to interpret and model a wealth of excited--state
phenomena in semiconductors.  Beside originating sub--band gap signatures in optical spectra, \optes have also been predicted to condensate, diffuse, recombine,
relax.  However, all these phenomena are rooted on a theoretical definition of the excitonic state based on the following simple picture: ``excitons'' are
actual particles that both appear as peaks in the linear absorption spectrum and also behave as well--defined quasiparticles.  In this paper we show, instead,
that the electron--phonon interaction decomposes the initial \textit{optical} (i.e., ``reducible'') excitons into \textit{elemental} (i.e., ``irreducible'') excitons,
the latter being a different kind of bound electron--hole pairs lacking the effect caused by the induced, classical, electric field.  This is demonstrated within
a real--time, many--body perturbation theory approach starting from the interacting electronic Hamiltonian including both electron--phonon and electron--hole interactions.  
We then apply the results on two realistic and paradigmatic systems, monolayer MoS$_2$ (where the lowest--bound \opte is optically inactive) and monolayer MoSe$_2$ (where it is optically active), using first--principles methods to compute the exciton--phonon coupling matrix elements.
Among the consequences of optical--elemental decomposition, we point to a homogeneous broadening of absorption peaks occurring even for the lowest--bound \opte, and we demonstrate this by computing exciton-phonon transition rates.
More generally, our findings suggest that the \optes gradually lose their initial structure and evolve as \undes. These states can be regarded as the real
intrinsic excitations of the interacting system, the ones that survive when the external perturbation and the induced electric fields have vanished.
\end{abstract}

\maketitle

\section{Introduction}\label{s:int}
%%%%%%%%%%%%%%%%%%%%%%%%%%%%%%%%%%%%%%%%%%%%%%%%%%%%%%%%%%%%%%%%%%%%%%%%%%%%%%%%%%%%%%%%%%%%%%%%%%%%%%%%%%%%%%%%%%%%%%%%%%%%%%
The exciton concept is crucial to contemporary condensed matter physics and materials science, since it allows for a simple description of the response of the
electrons in a crystal to an external electromagnetic field.  In fact, the usual interpretation of spectroscopic experiments states that, when an electron--hole
pair excitation is created in a (semiconducting) material, the Coulomb interaction will bind the pair creating an {\em exciton}\cite{Onida2002}.

{\em Excitons and optical spectra}. 
%----------------------------------
The excitonic picture stems from the interpretation of optical absorption spectra.
Absorption is linked to the difference between the external field and the total field inside the material, including the contribution coming from the induced
macroscopic polarization~\cite{Strinati1988}.
In fact, the exciton energies are just the frequencies of the time oscillation of the induced polarization~\cite{Attaccalite2011b}, which in turn is determined by electronic charge oscillations induced by the
external field. 
From a theoretical point of view, the excitonic picture is based on a combination of linear response theory 
--- involving weak external fields --- and many--body perturbation theory.
In the usual treatment this leads to the well--known and widely used Bethe--Salpeter equation\,(BSE)\cite{Strinati1988,Onida2002}.
The computational application of the BSE to a variety of realistic semiconducting
systems has led to crucial advances in the field of theoretical optical spectroscopy\cite{Bechstedt2015,Martin2016}.
In this paper, we will refer to excitations created by the interaction with an external electric field as \opte for the sake of 
clarity\footnote{In many--body theory they
correspond to the poles of the so--called reducible response function\cite{Strinati1988}.  They are also called ``singlet'' or ``transverse''
excitons\cite{Ambegaokar1960,DelSole1984} depending on the context.}.
When an \opte\, appears inside the electronic band gap of a material, the exciton is said to be bound (the lowest-bound, optically active exciton thus defines the \textit{optical} gap of the system).
The strength of the binding depends on several factors but, in general, it is stronger in
systems with wide optical gaps and low dimensionality due to weak screening of the electron--hole interaction\cite{Thygesen_2017}.
Within linear response theory, excitonic properties are computed in practice via the Hamiltonian representation of the BSE~\cite{Onida2002}. When retardation effects~\cite{Marini2003}  and electron--phonon interaction 
are neglected it is in fact possible to rewrite the excitonic state as an eigenstate of a pseudo--Hermitian matrix\cite{Gruuning2009,Martin2016}. 

{\em Excitons as real particles?} 
%----------------------------------------
Despite the fact that the pseudo--Hermitian structure of the BSE does not necessarily ensure that the exciton can be represented as a {\em real} 
bosonic particle\cite{pseudo-H}, the possibility that these charge oscillations also correspond to real populations of nonequilibrium bound electron--hole pairs is highly debated.
The question is how the interplay of the electromagnetic properties and internal structure of excitons may produce a real population and what is the role of the electron--phonon interaction in this process.
In a real--particle picture, \optes are treated as bosons weakly
coupled by an effective interaction.  
In the last few years several
theoretical~\cite{Moody2015,Selig2016,Christiansen2019} and experimental~\cite{Trovatello2020,Dong2021,Madeo2020,Dendzik2020} works have been using the real--particle assumption for \optes. 
The advent of ultra--fast physics has made it possible to investigate in real time the dynamics of photo--excited materials, increasing the interest in excitonic
physics. Model calculations~\cite{Koch2006,Katsch2018,Katsch2018} have boosted the concept of excitons as real particles, providing intuitive interpretations of the 
out--of--equilibrium experiments. 
Within this picture, \optes\, have been proposed to form, diffuse, relax, scatter and even condensate before
recombining\cite{Thranhardt2000,Kira2006,Selig2016,Christiansen2019,Miller2019,Tanimura2019,Perfetto2019,Ostreich1993}.  
Also using linear--response inspired model Hamiltonians, excitonic features have been predicted to appear in time--resolved ARPES~\cite{Rustagi2018,Christiansen2019,Kemper2020,Dong2021}.  
This approach has also been used to formulate an excitonic version of the semiconductor Bloch equation\cite{Haug2008} with the aim of modelling exciton dynamics in transition metal dichalcogenides (TMDs)\cite{Selig2016,Christiansen2019,Ovesen2019,Katsch2020}. 
A relevant feature of this approach is that excitons are treated in the popular Wannier model\cite{Mahan1990,Wannier1937,Elliott1957}, which describes excitons as a hydrogen--like energy level series stemming from parabolic electronic bands and having themselves a parabolic dispersion in reciprocal space.  
These studies cemented the very intuitive picture that after photo--excitation, the system can be described in terms of bound electron--hole pairs which are
essentially the same as the ones observed in optical absorption. 

The consensus on the \opte--as--particle picture  is, however, not complete. 
In the paradigmatic case
of MoS$_2$, for example, it was initially suggested that the observed rapid raise of the transient absorption signal was due to the ultra--fast formation of
excitons\cite{Trovatello2020}, supported by model calculations based on the excitonic Bloch equations\cite{Katsch2018}. Smejkal and al.\cite{Smejkal2021},
however, interpreted the very same experimental results purely in terms of single--body charge migrations and using \textit{ab initio} methods.
Additionally, it has also been shown that a perfect bosonization of interacting electron--hole pairs is impossible due to their fermionic substructure retaining the usual indistinguishability and Pauli repulsion properties\cite{Combescot2002,Combescot2004,Combescot2007}.
Finally, a natural consequence of treating \optes as real particles is that the lowest energy \opte must have an infinite lifetime (i.e., vanishing peak linewidth) since energy conservation does not allow any scattering. The same situation occurs in the electronic quasiparticle theory where the electronic linewidth is known to go to zero at the Fermi level.

{\em The Exciton-Phonon coupling}. 
%---------------------------------
The enormous interest in the excitonic dynamics has made it crucial to investigate the problem of exciton--phonon coupling.  In this case, the main conceptual
approach is again to consider excitons as bosons described by the BSE Hamiltonian.  Indeed, motivated by the pioneering works of Toyozawa, Segall and
Mahan\cite{Toyozawa1958,Segall1968,Toyozawa2003}, several authors have extended and upgraded the original model using many--body theory with the aim of
performing fully first--principles simulations on realistic materials, focusing on exciton relaxation lifetimes\cite{Chen2020}, spectral
functions\cite{Antonius2022} and exciton--phonon sidebands\cite{Cudazzo2020a,Cudazzo2020b}.
%In particular, Ref. [\onlinecite{Cudazzo2020b}] shows how the
%bosonic exciton model may actually be reobtained with less assumptions by including a first--order correction to the BSE kernel which involves the dynamical
%phonon propagator.

If the \opte\, is assumed to be a well--defined, boson--like particle, then the phonons can mediate the exciton--exciton interaction causing, for example, the
dressing of the excitons and finite lifetimes~\cite{Moody2015,Baldini2019}.  This picture has been employed to describe phonon--assisted sidepeaks in absorption
and luminescence spectra,\cite{Segall1968,Rudin1990,Perebeinos2005,Cannuccia2019,Paleari2019,Cudazzo2020b,Antonius2022} as well as the linewidths of exciton
peaks\cite{Toyozawa1958,Selig2016,Reichardt2020,Chen2020}.  The methodologies employed in the above references range from parametrised simple models to
tight--binding treatments and fully first--principles descriptions.  

\begin{figure}[t!]
  {\centering
  \includegraphics[width=\columnwidth]{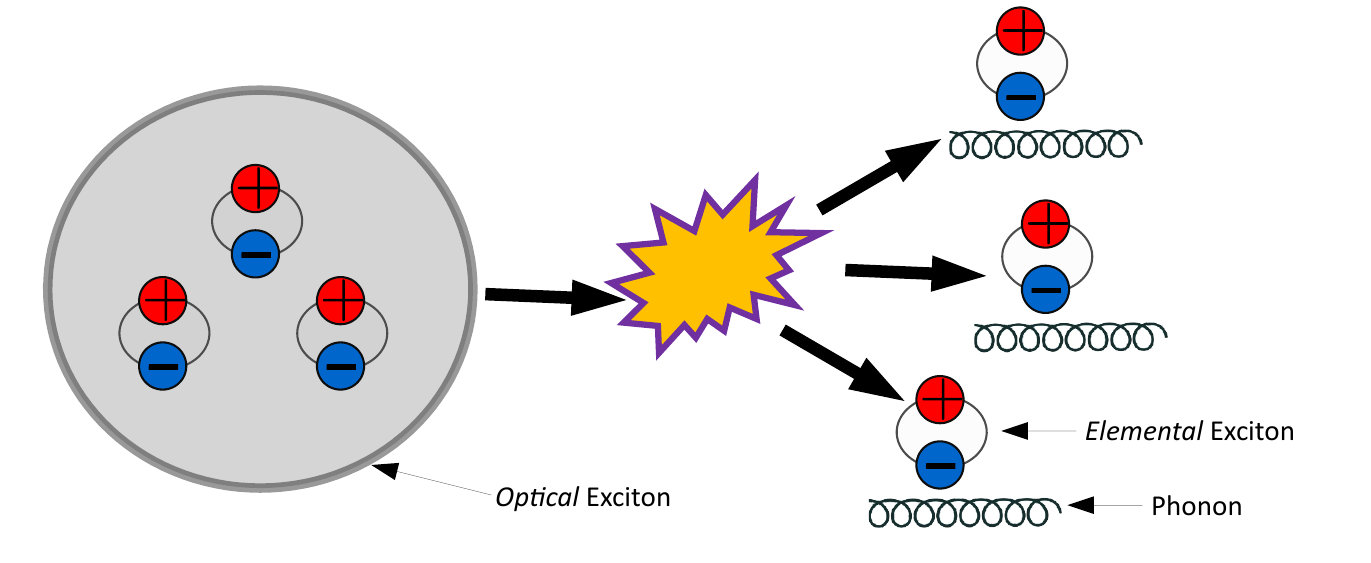}}
  \caption{Simple exciton--phonon coupling sketch. The picture we propose in this work is that the optical excitation is decomposed into a distribution of
\undes -- undressed of the Hartree interaction (i.e., the electron--hole exchange)  ---  by electron--phonon interactions. This distribution is the real internal structure of the
\opte. A remarkable consequence of this picture is that the lowest energy \opte\, has zero width in a real--particle representation, while if written in terms of
\undes acquires a finite, observable energy broadening.}
\label{fig:sketch} \end{figure}

{\em Revisiting the exciton--phonon picture}.
%--------------------------------
Despite its success, the treatment of excitons as real particles remains an underinvestigated \textit{assumption}.  
Thus, in this work we ask the following question:
given that ``excitons'' and ``phonons'' are both excitations dressed by the same electron--electron interactions, is it always sound to treat them as
``pristine'' particles that may interact with each other?  Or, rather, a proper account of their internal structures, consistent with the approximations we
generally use to treat electronic interactions, may lead to a subtler picture?  We aim to address this issue by considering the shaping of the exciton complex
by both external light and lattice vibrations.  To this end, we derive a theory describing the scattering of \optes\, with phonons.  In deriving our theory we
demonstrate that the \optes\, are scattered by the phonons in \textit{``elemental'' excitons}, undressed of the electron--hole exchange components, thus complicating the original simpler
picture.\footnote{In this paper we use the term \undes for simplicity. These states correspond, in many--body theory, to the poles of the \textit{irreducible}
response function and hence may also be called irreducible excitons. They also correspond to the ``triplet''\cite{Marsili2021} or ``Ising''
excitons\cite{Guo2019} in spin--polarised systems.} A graphical summary of this statement is presented in Fig. \ref{fig:sketch}.  A possible consequence is that
\optes, as defined in photoabsorption, may not provide a suitable basis to describe excited--state dynamics and excitonic lifetimes, as well as to calculate
phonon--assisted optical properties, especially in materials where electron--hole exchange interaction is large with respect to the excitonic binding energy. 
In order to explore this possibility, we perform fully first--principles exciton--phonon numerical simulations in monolayer MoS$_2$ and MoSe$_2$ with the goal
of comparing the standard ``real--particle'' approach with the one proposed here.  In particular, we show how \optes\, are decomposed in a packet including a large number
of \undes and how the homogeneous linewdith of even the lowest--bound \opte\, peak may be non--vanishing, according to our picture of exciton--phonon interaction. We provide a lower--bound estimate for these linewidths.

The paper is organised as follows. In Sec. \ref{s:excitons_intro} we introduce the many--body electronic Hamiltonian and the optical response function
describing absorption, and in Sec. \ref{s:GBSE} we derive a generalised Bethe--Salpeter equation to account for ``optical'' or ``elemental'' excitonic
properties. This is followed in Sec. \ref{s:optical_dress} by a discussion of the difference of the two pictures with \textit{ab initio} results for monolayer
MoS$_2$ and MoSe$_2$.  The exciton--phonon coupling problem is introduced and worked out theoretically in Sec. \ref{s:theo}.
Computational results for exciton--phonon coupling matrix elements and linewidths involving \textit{optical} and \textit{elemental} excitons in \mos\, and
\mose\, are reported in Sec. \ref{s:sim}, which is followed by a discussion in Sec. \ref{s:discuss}. 
The main text is complemented by four Appendices, including one reporting the full computational details.

\section{External fields and the exciton definition}\label{s:excitons_intro}
%%%%%%%%%%%%%%%%%%%%%%%%%%%%%%%%%%%%%%%%%%%%%%%%%%%%%%%%%%%%%%%%%%%%%%%%%%%%%%%%%%%%%%%%%%%%%%%%%%%%%%%%%%%%%%%%%%%%%%%%%%%%%%
The state--of--the art definition of \opte is based on the linear response of the interacting electron density to an external electromagnetic perturbation,
which is described by the Bethe--Salpeter equation, used in standard first--principles calculation of optical absorption
spectra\cite{Onida2002,Strinati1988,Martin2016}.  As before, we will refer to this equation as \OBSE, in order to keep it distinguished from the \EBSE
describing the elemental excitons.
%The \OBSE and \EBSE are derived in the Supporting  Informations\,(SI). 

We follow the real--time approach of Refs.~\cite{Perfetto2015,Attaccalite2011,Pal2009} in which our subsequent extension including dynamical electron--phonon interactions will also be formulated.
Let us consider the Hamiltonian of the electronic system perturbed by a scalar external potential $U_{ext}$.
We assume a purely longitudinal gauge with no external vector potentials.
\seq{
 \lab{e:h1}
\eqg{
 \h{H}\pr{t}=\h{H}_e+\int d\rr \h{\gr}\pr{\rr} U^{ext}\pr{\rr,t},\\
 \h{H}_e=\sum_i\h{h}_i+\h{W}_{e-e}+\h{V}_{e-ion}.
}
}
In \elab{e:h1}{a} $\h{H}_e$ is the electronic Hamiltonian which includes the electron--electron\,($\h{W}_{e-e}$) and the {\em bare} electron--ion\,($\h{V}_{e-ion}$)
interactions. $U^{ext}$ represents the {\em total} time--dependent perturbation, which embodies the experimental field and the macroscopic part of the field induced
in the material.

It is essential to note that, in \e{e:h1}, $\h{V}_{e-ion}$ is not assumed to be screened from the beginning, as done in the previous
works~\cite{Antonius2022,Chen2020}. It has been, indeed, demonstrated~\cite{Marini2015} that such an assumption implies double--counting problems that
can be avoided only by screening dynamically $\h{V}_{e-ion}$ along with the solution of the BSE.

In \e{e:h1} we have also introduced the electronic density operator $\h{\gr}\pr{\rr}$ and the single--particle Hamiltonian $\h{h}_i$.
Let us also mention the electron Green's function $G(\rr_1 t_1,\rr_2 t_2)$ associated with this Hamiltonian and recall that the density is given by $\rho(\rr
t)=\average{\h{\gr}\pr{\rr}}=-\im G(\rr t, \rr^+ t^+)$.  
By using diagrammatic methods the effect of $\h{W}_{e-e}$ and $\h{V}_{e-ion}$ is translated in a self--energy potential $\Sigma$ which appears in the equation
of motion for $G$, as we will discuss shortly. $\Sigma$ is comprised of the classical Hartree interaction,
\eq{
 \lab{e:Vh}
 V^H\(\rr,t\) = \int^\prime \di \rr^\p \rho\(\rr^\p,t\) v\(\rr,\rr^\p\),
}
where $v\(\rr,\rr^\p\)$ is the bare Coulomb potential, plus an exchange and correlation part treatable at different levels of approximation~\cite{Attaccalite2011b}.
The induced
potential is actually the macroscopic average of $V^H$ and we include it by definition in $U^{ext}$.
Thus, the integral in \e{e:Vh} runs only on the spatial microscopic
components. More details about this choice are provided in \app{APP:fields}.

The variation of the density with respect
to the external field within linear order defines the electronic, optical response function $\chi_{opt}$:
\eq{
\lab{e:chi1}
\chi_{opt}(\rr_1 t_1,\rr_2 t_2)=\left.\frac{\delta \rho(\rr_1,t_1)}{\delta U^{ext}(\rr_2,t_2)}\right|_{U^{ext}=0}.
}
Optical excitons are the poles of the Fourier transform of $\chi_{opt}$ with respect to the time difference.
A key property of \e{e:chi1} is the appearance in the denominator of $U^{ext}$, which is a macroscopic field and the one that can be experimentally observed. This has a crucial impact on the definition of the \optes. 

Let us now consider a hypothetical experimental apparatus able to also detect all the microscopic variations, induced by light
absorption, of the \textit{total} potential $U^{tot}=U^{ext}+V^{H}$. 
This corresponds to having an electric field detector with a spatial resolution tinier than the unit cell size of the system.
As a consequence, in this case, the experimental observable would be described by a different response function, that we denote {\em elemental} and that is defined as:
\eq{
\lab{e:chi_el}
\chi^{el}\(\rr_1 t_1,\rr_2 t_2\)=\left.\frac{\delta \rho\(\rr_1,t_1\)}{\delta U^{tot}\(\rr_2,t_2\)}\right|_{U^{ext}=0}.
}
\e{e:chi1} and \e{e:chi_el} emphasize that the definition of {\em exciton} as an observable is determined by the measure process. 
%As in the system also microscopic potentials
%appear it is natural to ask what is the right definition of the response function defining the microscopic excitonic states. 

\section{The generalized Bethe--Salpeter equation}\label{s:GBSE}
%%%%%%%%%%%%%%%%%%%%%%%%%%%%%%%%%%%%%%%%%%%%%%%%%%%%%%%%%%%%%%%%%%%%%%%%%%%%%%%%%%%%%%%%%%%%%%%%%%%%%%%%%%%%%%%%%%%%%%%%%%%%%%
Given the definitions \e{e:chi1} and \e{e:chi_el} the corresponding \textit{optical} and \textit{elemental} BSEs can be easily derived by neglecting the electron--phonon interaction and using the 
non--local Hartree plus screened exchange (HSEX) scheme\cite{Onida2002}  in which $\Sigma$ is written as the sum of the classical, mean--field Hartree term $\h{V}_H$ with the statically
screened exchange interaction embodied in the ``mass'' term $\h{M}_{SEX}$.
\eq{
 \lab{e:hsex1}
 M^{SEX}\(\rr_1,\rr_2,t\)=\im G\(\rr_1 t,\rr_2 t^+\)W\(\rr_1,\rr_2\).
}
Here $W=\epsilon^{-1}_H v$ is the screened Coulomb interaction, and the static screening $\epsilon^{-1}_{H}$ is
calculated in the Hartree (also called random phase, RPA) approximation\cite{Onida2002}.
%In order to proceed, in the following section we provide the derivation of a generalized form of the BSE, and we introduce the formalism
%we will keep using also in the exciton--phonon case.

In the following we will first introduce a convenient basis to write the Dyson equation. We will then derive from the Hedin's representation of the mass operator a
generalized Bethe--Salpeter equation for the three--points electron--hole propagator. From this general equation we will derive the \OBSE and \EBSE.
Equipped with the two Bethe Salpeter equations, we resume our discussion of optical vs elemental excitons --- leading to the problem of exciton--phonon coupling --- in Section \ref{s:optical_dress}.

\subsection{Dyson's equation in a generalized basis}\lab{ss:hedins}
%=============================================================================================================================
The electronic Green's function (GF) $G\(\rr_1 t_1,\rr_2 t_2\)$ corresponding to the Hamiltonian \e{e:h1} satisfies the Dyson equation for the single--particle
Green's function\cite{Stefanucci2013}
\begin{widetext}
\eq{
\lab{e:T1}
 G\(\rr_1 t_1,\rr_2 t_2\)=
\int d\rr_3 \rr_4 d t_3 t_4 G^0\(\rr_1 t_1,\rr_3 t_3\) \left[\gd\ttw{2}{3}\gd\(\rr_2,\rr_3\)+\gS\(\rr_3 t_3,\rr_4 t_4\) G\(\rr_4 t_4,\rr_2 t_2\)\right].
}
\end{widetext}
The {\bf exact} self--energy $\gS$ corresponding to \e{e:h1} has been derived, among others, in Ref. [\onlinecite{VanLeeuwen2004}]:
\ml{
\lab{e:T2a}
\gS\(\rr_1 t_1,\rr_2 t_2\)=M\(\rr_1 t_1,\rr_2 t_2\)+\\+V^{H}\(\rr_1,t_1\)\gd\(t_1,t_2^+\)\gd\(\rr_1,\rr_2\),
}
with
\eq{
 \lab{e:T2}
M\(\rr_1 t_1,\rr_2 t_2\)=M^{e-e}\(\rr_1 t_1,\rr_2 t_2\)+M^{e-p}\(\rr_1 t_1,\rr_2 t_2\).
}
In \e{e:T2} $M^{e-e}$ and $M^{e-p}$ are, respectively, the electron--electron (e-e) and the electron--phonon (e-p) terms. 
The approximated mass operator used in this work will be described later. We first want to derive some exact properties of the response function.
We start by introducing a convenient single--particle representation:
\eq{
 \lab{e:T3}
 G\(\rr_1 t_1,\rr_2 t_2\)=\sum_{ij} \oo{\phi}_i\(\rr_1\) \phi_j\(\rr_2\) G_{ij}\ttw{1}{2},
}
with $\{\phi_i\(\rr\)\}$ a suitable complete basis\,($i$ represents a generic electronic band and $\kk$--point). Thanks to \e{e:T3} we can rewrite \e{e:T1} in a compact form using a matrix notation
\eq{
\lab{e:T4}
 \uu{G}\ttw{1}{2}=\uu{G}^0\ttw{1}{3}\[\gd\ttw{3}{2}+\uu{\gS}\ttw{3}{4}\uu{G}\ttw{4}{2}\].
}
Quantities that depend on two electronic indices are represented as matrices $\[\uu{O}\]_{ij}$. In the following, more convoluted objects
depending on four indices will appear. In this case we will represent them as tensors: $\[\,\uu{\uu{O}}\,\]\mind{ij}{kl}$. 
The conventions used to represent tensorial operations are defined in \app{APP:conventions}. The Einstein convention (assuming all
repeated indices to be summed) is also implied. 
\subsection{The Generalized Bethe--Salpeter Equation}\lab{ss:GBSE}
%=============================================================================================================================
In the single--particle basis representation all response functions are tensors of rank $2$. In particular we can define a generalized two--particle Green's function
\eq{
 \lab{e:G1}
 L^{\eta}\mind{ij}{kl}\tth{1}{2}{3}\equiv \frac{\gd G_{ij}\ttw{1}{2}}{\gd \eta_{kl} \(t_3\)}.
}
\e{e:G1} is easily connected to the optical/elemental response functions. Indeed, by definition
\eq{
 \lab{e:G1.1}
 \uu{\uu{\chi}}^{opt/el}\ttw{1}{2}\equiv -i\left. \uu{\uu{L}}^{\eta}\(t_1,t_1^+;t_2\)\right|_{\eta=U^{ext}/U^{tot}}.
}
Therefore in \e{e:G1} $\eta$ is an arbitrary field that, later, we will assume to correspond to $U^{tot}$ or $U^{ext}$ .
We can work out \e{e:G1} by  differentiating \e{e:T3} 
\ml{
 \lab{e:G2}
 \frac{ \gd \uu{G}\ttw{1}{2} } {\gd \uu{\eta}\(t_3\)}= 
   \uu{G}^0\ttw{1}{4}\left[\uu{\gS}\ttw{4}{5} \uu{\uu{L}}^{\eta}\tth{5}{2}{3}+\right. \\
   \left.+ \frac{\gd\uu{\gS}\ttw{4}{5}}{\gd \uu{\eta}\(t_3\)} \uu{G}\ttw{5}{2} \right].
}
We now need to calculate $\frac{\gd\uu{\gS}\ttw{4}{5}}{\gd \uu{\eta}\(t_3\)}$. By using the Dyson equation and the chain rule we get
\eq{
 \lab{e:G4}
 \frac{\gd\uu{\gS}\ttw{4}{5}}{\gd \uu{\eta}\(t_3\)}=\uu{G}^{-1}\ttw{4}{6} \uu{\uu{L}}^{\eta}\tth{6}{7}{3} \uu{G}^{-1}\ttw{7}{5}.
}
We see that the equation for $L^{\eta}$ can be closed:
\begin{widetext}
\ml{
 \lab{e:G5}
 \uu{\uu{L}}^{\eta}\tth{1}{2}{3}=\uu{G}\ttw{1}{3} \uu{\uu{\gd}}\, \uu{G}\ttw{3}{2}+\\+
  \uu{G}\ttw{1}{4}\[\uu{\uu{\Xi}}\tfo{4}{5}{6}{7}+\uu{\uu{K}}^{\eta}\ttw{4}{6}\gd\ttw{4}{5}\gd\ttw{6}{7}\]
  \uu{\uu{L}}^{\eta}\tth{6}{7}{3} \uu{G}\ttw{5}{2}.
}
\end{widetext}

In \e{e:G5} we have introduced the tensorial delta function $\[\uu{\uu{\gd}}\]\mind{ij}{kl}=\gd_{ik}\gd_{jl}$ and
\seq{
 \lab{e:G6}
\eqg{
 \uu{\uu{\Xi}}\tfo{4}{5}{6}{7}\equiv   \frac{\gd\uu{M}\ttw{4}{5}}{\gd \uu{G}\ttw{6}{7}},\\
 \uu{\uu{K}}^{\eta}\ttw{4}{6}\equiv   \frac{\gd\( \uu{U}^{tot}\(t_4\)-\uu{\eta}\(t_4\)\)}{\gd \uu{G}\ttw{6}{6^+}}.
}
}
\e{e:G5} represents the {\em generalized BSE} and it allows to connect any self--energy $\gS$ to the equation of motion for the two--particles green's function
$L$. 

\subsection{The {\em optical} and {\em elemental} Bethe--Salpeter Equations}\lab{ss:OBSE_EBSE}
%=============================================================================================================================
The \textit{optical} (or reducible) BSE is obtained when $\eta\(\rr,t\)\equiv U^{ext}\(\rr,t\)$ and the electronic self--energy is approximated with the SEX expression neglecting
electron--phonon effects. In this
case
\EQ{
 \lab{e:S1}
 M^{e-e}_{ij}\ttw{1}{2}\approx i \gd\(t_2,t_1^+\) W_{\substack{ij\\kl}} G_{kl}\(t_1,t_1^+\),
}
Where we defined
\eq{
 \lab{e:S3}
 W_{\substack{ij\\kl}}\equiv \int d\rr_1 \rr_2 \oo{\phi}_i\(\rr_1\) \phi_k\(\rr_1\) W\(\rr_1,\rr_2\) \oo{\phi}_l\(\rr_2\) \phi_j\(\rr_2\).
}
It follows that
\seq{
 \lab{e:S4}
\eqg{
 K^{opt}\ttw{4}{6}\mind{ij}{kl}=-i V^{H}\mind{ik}{jl}\gd\(t_4,t_6\),\\
 \Xi\tfo{4}{5}{6}{7}\mind{ij}{kl}=i\gd\ttw{1}{2}\gd\ttw{4}{3^+}W\mind{ij}{kl}.
}}
Here the repulsive term $K^{opt}$ is determined by the microscopic components of the Hartree interaction, i.e., by the local field effects.
This term is also known as electron--hole exchange because it swaps electron and hole indices with respect to $\Xi$, which is known instead as the direct or binding term.
In this case \e{e:G5} is closed in the space of two times, $\uu{\uu{L}}^{opt}\ttw{1}{2}\equiv \uu{\uu{L}}\(t_1,t_1^+;t_2\)$, and
\ml{
 \lab{e:S5}
 \uu{\uu{L}}^{opt}\ttw{1}{2}=\uu{G}\ttw{1}{2} \uu{\uu{\gd}}\, \uu{G}\ttw{2}{1}+\\
  +i \uu{G}\ttw{1}{3}\[\uu{\uu{W}}-\uu{\uu{V}}^{H}\] \uu{\uu{L}}^{opt}\ttw{3}{2} \uu{G}\ttw{3}{1}.
}
The same procedure can be applied to derive the \textit{elemental} (or irreducible) BSE which corresponds to taking $\eta\(\rr,t\)\equiv U^{tot}\(\rr,t\)$, from
which follows $K^{el}=0$. In this case, then, we have
\ml{
 \lab{e:S6}
 \uu{\uu{L}}^{el}\ttw{1}{2}=\uu{G}\ttw{1}{2} \uu{\uu{\gd}}\, \uu{G}\ttw{2}{1}+\\
 +i \uu{G}\ttw{1}{3}\uu{\uu{W}}\, \uu{\uu{L}}^{el}\ttw{3}{2} \uu{G}\ttw{3}{1}.
}
The last step we need is to connect the $L^{opt/el}$ to the diagonalization of the Bethe--Salpeter Hamiltonian, which is used in practice to compute exciton
energies and wave functions. The procedure is outlined in \app{APP:BS_hamiltonian} and leads to the definition of
\EQ{
\lab{e:HOBSE}
  \mc{H}^{opt}\mind{ij}{kl}=\gd_{ik}\gd_{jl}\(\gee_i-\gee_j\)-\(f_j-f_i\)\(W\mind{ij}{kl}-K^{opt}\mind{ij}{kl}\).
}
Here again we see that, in the case of \optes, the repulsive term $K^{opt}$ is given by the microscopic part of the total field inside the material.
It is this purely electrostatic contribution which defines the \opte.

The two--particles Green's function relative to \undes has the same form as above, but now only $W$ appears in the definition of the excitonic Hamiltonian 
$\mc{H}^{el}$:
\eq{
\lab{e:HEBSE}
  \mc{H}^{el}\mind{ij}{kl}=\gd_{ik}\gd_{jl}\(\gee_i-\gee_j\)-\(f_j-f_i\)W\mind{ij}{kl}.
}
It is essential to observe that both $\uu{\uu{W}}$ and $\uu{\uu{K}}^{opt}$ are, in general, pseudo--Hermitian matrices~\cite{Gruuning2009} and, therefore, even if their eigenvalues
are real the left and right eigenvectors are different~\cite{pseudo-H}. We assume here that $\mc{H}^{el/opt}$ are strictly Hermitian, so that
once diagonalized we can finally write $\uu{\uu{L}}^{opt/el}$ in an excitonic representation.  By calling $E^{opt/el}_{\gl}$ the eigenvalues of $\uu{\uu{\mc{H}}}^{opt/el}$, we have that \seq{
\lab{e:OEBSE.freq}
\eqg{
  L^{opt}\(\go\)=\sum_{\gl} \frac{1}{\go+i0^+-E^{opt}_{\gl}},\\
  L^{el}\(\go\)=\sum_{\gl} \frac{1}{\go+i0^+-E^{el}_{\gl}}.
}}
The picture that follows from the Hamiltonian representation of the BSE is simple: an {\em optical}/{\em elemental} exciton is a superposition of electron--hole
pairs weighted by the eigenvectors of the excitonic Hamiltonian, whose components we call $\uu{A}_{\gl}^{opt/el}$:
\eq{
\lab{e:OEBSE.ket}
 \ket{\gl^{opt/el}}=\sum_{ij}A_\gl^{ij,opt/el} \ket{i}\otimes\ket{j}.
}

\section{Absorption spectra: {\em optical} and {\em elemental} excitons}\label{s:optical_dress}
%%%%%%%%%%%%%%%%%%%%%%%%%%%%%%%%%%%%%%%%%%%%%%%%%%%%%%%%%%%%%%%%%%%%%%%%%%%%%%%%%%%%%%%%%%%%%%%%%%%%%%%%%%%%%%%%%%%%%%%%%%%%%%
\begin{figure*}%[hbtp]
  {\centering \includegraphics[width=\textwidth]{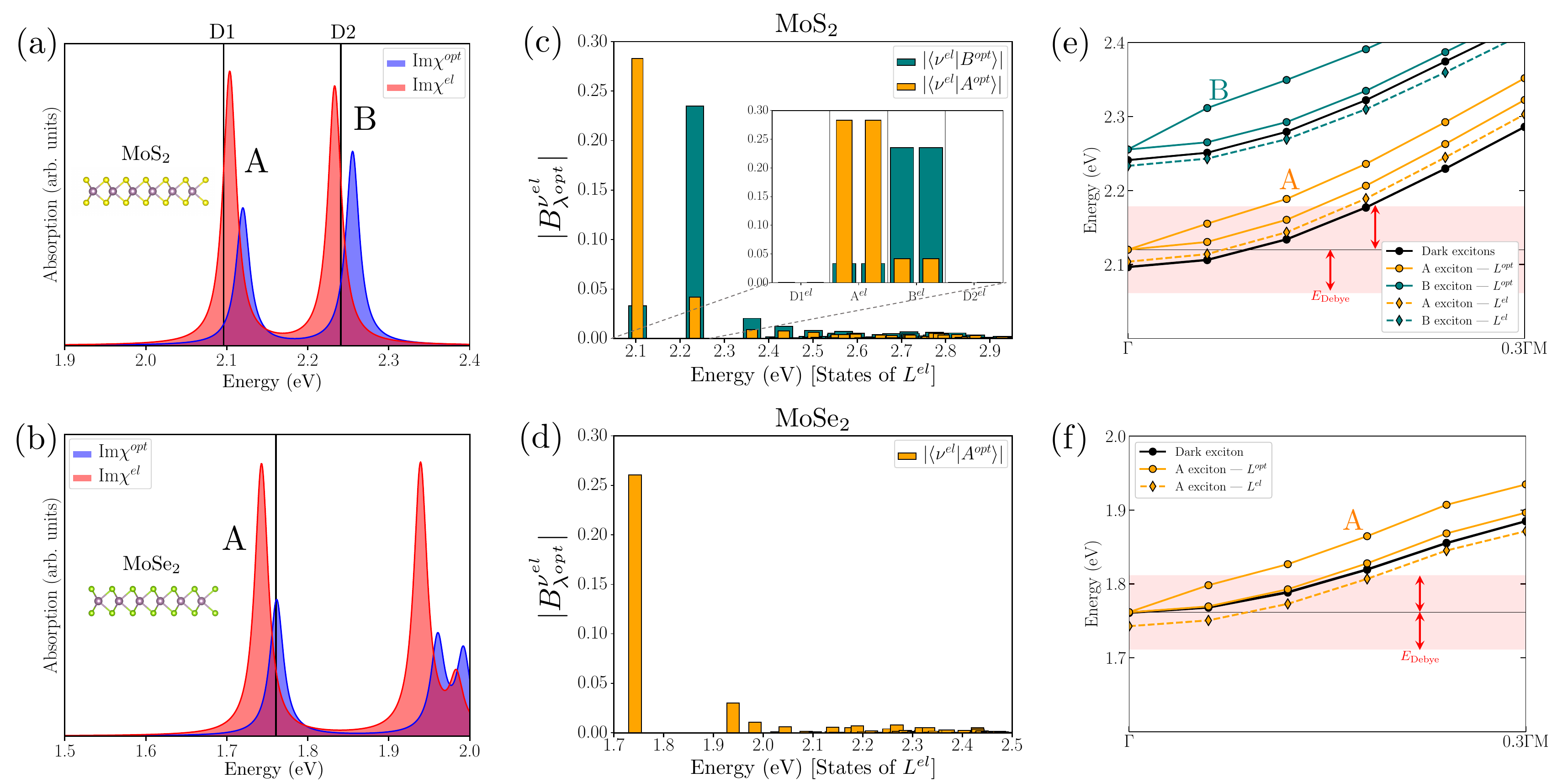}}
\caption{{\em Optical} and {\em elemental} excitons compared in the case of monolayer MoS$_2$ (upper panels) and monolayer MoSe$_2$ (lower panels). Panels (a--b): 
\textit{Optical} (blue) and \textit{elemental} (red) spectrum. The black vertical lines
represent the dark states, while the peak ``A'' and ``B'' labels represent the established names for the relative bright peaks in the literature. In the insets showing
the lattice geometry the gray balls represent Mo atoms, the yellow (green) ones represent S (Se) atoms. A set broadening parameter of $10$ meV was used for the peaks.
Panels (c--d): Projections (normalised) of A and B \optes onto the \undes (see \e{eq:B_proj}). The projections of the A (B) optical
excitons are shown in orange (teal). The inset shows a zoom around the low--lying dark (D$1$, D$2$) and bright (A, B) states of MoS$_2$ (they are all doubly
degenerate).
Panels (e--f): Exciton dispersion curves. Full lines and circles: \textit{optical} excitons (including
$K^{opt}$--driven degeneracy splittings). Dashed lines and diamonds: \textit{elemental} excitons. The orange (teal) lines track the dispersion of
the A (B) excitons. Black lines follow the dispersion of the dark states. The red shaded regions marks the energy--allowed region for the exciton--phonon
scattering of the A exciton at $\Gamma$ (one Debye energy).
}\label{fig:EL_vs_OPT}
\end{figure*}

The differences between \optes and \undes\, states are analyzed in detail in \fig{fig:EL_vs_OPT} for two paradigmatic TMDs: monolayer MoS$_2$ (upper panels) and
monolayer MoSe$_2$ (lower panels). Naturally, 
%{ energies (i.e., the spectral peak positions) of $\chi^{el}$ and $\chi^{opt}$ 
the strength and role of the matrix elements of
$K^{opt}$ are of particular interest to us.  
Since this is a repulsive contribution, lowering the binding energy of the excitons, $\chi^{el}$ may have in general more tightly
bound excitons than $\chi^{opt}$. Furthermore, the strength of the excitonic coupling with light is also affected, since it depends on the exciton wave
functions.

Optical absorption is connected to the imaginary part of the macroscopic dielectric function $\epsilon_M$, which is given by the $q\rightarrow 0$ limit of the response function $\chi$\cite{Strinati1988,Martin2016}:
\eq{
\lab{e:opt_abs}
\mathrm{Im} \epsilon^\eta_M(\omega) \propto \mathrm{Im} \lim_{q\rightarrow 0}\chi^\eta(q,\omega) = \sum_\lambda | \sum_{ij} A_\lambda^{ij,\eta} d_{ij}|^2 \delta(\omega -E_\lambda^\eta).
} 
Here, $d_{ij}$ is the optical matrix element for the light--induced electronic transition between states $i$ and $j$, calculated in the dipole approximation.

In \figlab{fig:EL_vs_OPT}{a--b} the absorption spectra of the two kinds of excitons are shown.
%These important differences are shown in \fig{fig:schematic_ABS} for two paradigmatic TMDs: MoS$_2$ (a) and MoSe$_2$ (b). 
The red region corresponds to the
absorption from \undes, while the blue region from \optes.  Each bright exciton also has a dark companion, shown with a black vertical line, in which
$K^{opt}=0$ always because of the opposite spin polarisation of electron and hole.\cite{Deilmann2017,Robert2020,Arora2015b} Thus, dark states always coincide in
both the optical and the elemental cases.  Conversely, the energy of the bright states changes and shifts downwards. In particular, the B exciton in MoS$_2$ and
the A exciton in MoSe$_2$ both slide below the energy of their dark companions.  The latter case is notable because it means that in MoSe$_2$ the lowest \OBSE
state is dark, whereas the lowest \EBSE is bright.  The energy difference between the corresponding optical and elemental states are: $\Delta^{\mathrm{MoS}_2}A
= 16$ meV, $\Delta^{\mathrm{MoS}_2}B = 22$ meV and $\Delta^{\mathrm{MoSe}_2}A = 19$ meV.  Their intensity also noticeably changes, with $K^{opt}$ accounting for
a large increase of the B exciton with respect to the A one in MoS$_2$, while in the \textit{elemental} case they have almost the same intensity.\cite{Guo2019}
%It is then evident that the it is the kind of experimental probe which defines its corresponding absorption spectrum.

The relationship between optical and elemental excitons can be further elucidated by looking at the decomposition of one type into the other, defining the projections
\EQ{
\label{eq:B_proj}
B_{\lambda^{opt}}^{\nu^{el}}=\sum_{ij} \oo{A_\nu^{ij,el}} A_\gl^{ij,opt} = \braket{\nu^{el}}{\gl^{opt}}.
}
In  \figlab{fig:EL_vs_OPT}{c--d}  we report the calculated projections (values of $|B_{\lambda^{opt}}^{\nu^{el}}|$) of the lowest--bound \optes onto the \undes of
MoS$_2$ (c) and MoSe$_2$ (d).  In the case of  MoS$_2$ there are two bound states (A and B) while in  MoSe$_2$ only A.  Let us first look at the inset in Fig.
\fig{fig:EL_vs_OPT}(c); here we see that despite the A and B excitons of MoS$_2$ being formed by different eletronic transitions (the difference is due
to spin--orbit coupling), they both partially decompose onto each other: the A \opte has a sizable component onto the B \unde and viceversa.  This was already
noted in Ref. \onlinecite{Guo2019}: however, the decomposition shown in the inset only accounts for $30\%$ of the projection components, as can be evinced by
looking at the complete figure which includes up to $450$ \textit{elemental} excitonic states.  A large number of exchange--less excitonic states, much higher in energy than the A
and B ones, have nonzero projections with the optical A and B states, accounting for the remaining $70\%$ of the strength.  This suggests that a large
distribution of (elemental) states may play a role in the processes of optical--elemental scatterings and exciton dynamics.  The same is true in the case of the
A exciton in MoSe$_2$.

%\begin{figure}%[hbtp]
%  {\centering \includegraphics[width=\columnwidth]{TMD_results/Projections_Figure.pdf}}
%\caption{Projections (normalised) of A and B \optes onto the \undes (see \e{eq:B_proj}). (a) MoS$_2$, (b) MoSe$_2$. The projections of the A (B) optical
%excitons are shown in orange (teal). The inset shows a zoom around the low--lying dark (D$1$, D$2$) and bright (A, B) states of MoS$_2$ (they are all doubly
%degenerate).}\label{fig:elemental_excitons}
%\end{figure}

%This question would not be relevant for systems, like hexagonal boron nitride, where we expect \opte and \undes to be very similar.\cite{Paleari2018} However,
%for systems with a large Hartree contribution this is not true as shown in Fig. \ref{fig:exc_dispersion}.  
In \figlab{fig:EL_vs_OPT}{e--f} we report the
exciton dispersions obtained by solving the \OBSE (full lines) and \EBSE (dashed lines) at finite momenta.  These states, not observable with optical
light\cite{Madeo2020}, correspond to electronic transitions in which the electron momentum $k$ and the hole momentum $k^\prime$ differ as $k-k^\prime =q$.  The
plots are made close to $q=0$ ($\Gamma$) and along the direction $\Gamma M$ in the hexagonal, two--dimensional BZ of these systems.  
%Our previous Figure,
%\fig{fig:schematic_ABS}, corresponds to the states at $\Gamma$.  
We see that the bright A (orange color) and B (teal color) excitons, as well as their dark
companions (black color), are doubly degenerate states.  In particular, in addition to the energy shifts between optical and elemental bright excitons, we see
that the presence of the Hartree contribution in the \textit{optical} case causes a splitting of the bright excitons at finite momentum, something that is
completely absent in the \textit{elemental} case. Moreover, $K^{opt}$ causes the higher energy split state to have a linear behavior with respect to $|q|$ ---
something that is well--known in the literature\cite{Qiu2015,Deilmann2017} --- instead of the parabolic dispersion typical of $W$, which appears for all other
states.  If we consider for example the A excitons at $\Gamma$, we know that the scattering to finite--q states mediated by one phonon can take place in an
energy window with the size of the Debye energy ($59$ meV in MoS$_2$, $50$ meV in MoSe$_2$ according to our \textit{ab initio} calculation).  This window is
shown by the red shaded region: we can clearly see how, depending on the kinds of initial and final excitonic states to be considered in our exciton--phonon
description, the scattering dynamics may be quite different.  This is particularly relevant because the intraband, low--$q$ scattering mediated by acoustic
phonons is predicted to account for a large part of the excitonic homogeneous linewidths.\cite{Selig2017,Henriques2021}
%\begin{figure*}[hbtp]
%  {\centering \includegraphics[width=0.9\textwidth]{TMD_results/Dispersion_Figure.pdf}}
%  \caption{Calculated exciton dispersion curves for MoS$_2$ (a) and MoSe$_2$ (b). Full lines and circles: \textit{optical}/reducible dispersion (including
%$K^{opt}$--driven degeneracy splittings). Dashed lines and diamonds: \textit{elemental}/irreducible dispersion. The orange (teal) lines track the dispersion of
%the A (B) excitons. Black lines follow the dispersion of the dark states. The red shaded regions marks the energy--allowed region for the exciton--phonon
%scattering of the A exciton at $\Gamma$ (one Debye energy).} \label{fig:exc_dispersion}
%\end{figure*}

\section{The exciton--Phonon scattering}\label{s:theo}
%%%%%%%%%%%%%%%%%%%%%%%%%%%%%%%%%%%%%%%%%%%%%%%%%%%%%%%%%%%%%%%%%%%%%%%%%%%%%%%%%%%%%%%%%%%%%%%%%%%%%%%%%%%%%%%%%%%%%%%%%%%%%%
Several different theoretical approaches to the derivation of exciton--phonon coupling for computational purposes are available in the literature.\cite{Chen2020,Cudazzo2020b,Antonius2022}
All current approaches share the conceptual basis of the pioneering modellistic works of Toyozawa\cite{Toyozawa1958,Toyozawa2003} and Segall, Rudin and Mahan\cite{Segall1968,Rudin1990}.
%These works appeared before the first
%\ai\, calculations of excitonic effects in optical spectra\cite{Albrecht1998}, where the importance of the electrostatic contribution to the
%total \OBSE kernel was revealed.
%Indeed there are two main ingredients at the basis of the work of Toyozawa, Segall, Rudin and Mahan: \optes and \undes are assumed to be the same and the electron--phonon
This theory is based on three core assumptions: \optes and \undes are the same, the excitonic Hamiltonian representation is taken as granted
and the electron--phonon interaction appearing in the Hamiltonian is screened from the beginning. 

%As mentioned in the introduction it is well--known~\cite{Marini2003} that the  Hamiltonian representation, \e{e:HOBSE} and \e{e:HEBSE}, is valid only when the interaction $W$
%is assumed to be static. In addition, 
%%Those two assumptions, however, may not work well in realistic materials where the repulsive contribution in the kernel is large,
%as was discussed in the previous section,
%%. Indeed, especially 
%at low $q$--momenta close to the long wave--length limit $\chi^{el}$ and $\chi^{opt}$ can be very different. Finally,  it is also known  
%that the dressing of the electron--phonon interaction occurs dynamically and should not be introduced from the beginning~\cite{Marini2015}.

By assuming that only one kind of excitons $\eta$ exists, this theory also assumes $\mc{H}^{\eta}$ to be a physical Hamiltonian.
The eigenvectors of $\mc{H}^{\eta}$ are then used to define excitonic creation and annihilation operators
$\[\h{B}^{\eta}_{\gl}\]^\dagger$ and $\h{B}^{\eta}_{\gl}$, while $\mc{H}^{\eta}$ is rewritten as
\eq{
 \lab{e:Hbos.1}
 \h{\mc{H}}^{\eta} \approx  \sum_{\ga q}E^{\eta}_{\ga q}\[\h{B}^{\eta}_{\ga q}\]^\dagger\h{B}^{\eta}_{\ga q}.
}
An additional, crucial --- and strong\cite{Combescot2002,Combescot2004,Combescot2007} ---  assumption in \e{e:Hbos.1} is that excitons are good bosons in the sense that they satisfy
bosonic commutation relations. 

Now, by analogy with the electronic case, this bosonised treatment introduces exciton--phonon interaction into \e{e:Hbos.1} in the form
%We may thus postulate that exciton--phonon interaction arises from the linear coupling between excitons and phonon displacement
%operators. This leads to the following model Hamiltonian:
\eq{
\lab{e:Hbos.2}
\h{\mc{H}}^{\eta}_{e-p}=\sum_{\ga\ga^\p \mu q q^\p} \CG^{\mu q^\p}_{\ga \ga^\p q}\(\h{b}_{\mu q^\p}+\h{b}^\dagger_{\mu-q^\p}\)\(\h{B}^{\eta}_{\ga^\p q+q^\p}\)^\dagger\h{B}^{\eta}_{\ga q}.
}
This term represents the $\eta$--exciton--phonon interaction with $\h{b}_{\mu}$ being the phonon destruction operator for the phonon of branch $\mu$ and momentum $q^\p$.  The
crucial quantity in \e{e:Hbos.2} is the exciton--phonon coupling matrix element $\CG^{\mu q^\p}_{\ga \ga^\p q}$. 

Clearly \e{e:Hbos.2} makes sense only if it is possible to demonstrate that the case $\eta=opt$ involves only \optes. We will demonstrate in the next section
that this is, actually, not possible. 
Moreover a crucial consequence of \e{e:Hbos.2} is that the Fermi golden--rule predicts the excitonic linewidths, $\gamma_{\ga q}$, of the state $\ga$ with
momentum $\qq$ to have the form
\eq{
\lab{e:wrong_width}
\gamma_{\ga q} \propto \sum_{\mu\gb q^\p} |\CG_{\ga \gb q}^{\mu q^\p}|^2 \delta(E^{opt}_{\ga q} - E^{opt}_{\gb q+q^\p}\pm\Omega_{\mu q^\p}).
}
\e{e:wrong_width} has been used, for example, in Refs. [\onlinecite{Chen2020,Antonius2022}] and predicts the lowest \opte to have zero width.

Below, we will outline a derivation of the exciton--phonon coupling starting from the electronic Hamiltonian, \e{e:h1}, with electron--phonon interactions
explicitly included from the start. This will allow us to overcome several of the assumptions underpinning the ``bosonised'' excitons model.

\subsection{The vertex function}
\begin{figure}%[H]
\begin{center}
\parbox[c]{8cm}{
\begin{center}
(a)\\
\epsfig{figure=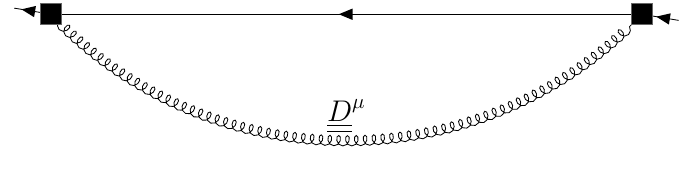,width=0.9\columnwidth}
\end{center}
}
\parbox[c]{8cm}{
\begin{center}
(b)\\
\epsfig{figure=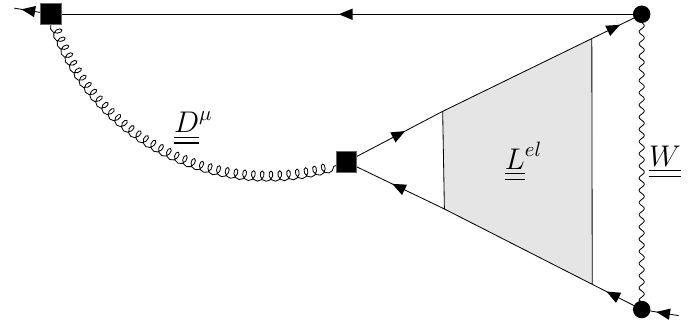,width=0.9\columnwidth}
\end{center}
}
\caption{Diagrammatic decomposition of the $M^{e-p}$ mass operator in terms of elemental electron--hole Green's function. (a) The well--known Fan--Migdal
self--energy ($M^{e-p}$ reduces to this when $\uu{\uu{\gC^{el}}}=\uu{\uu{\delta}}$). The straight (curly) line represents the electronic (phononic) GF and the
filled box ($\blacksquare$) is the dressed e-p coupling matrix element $\uu{g}$ from \e{e:g_dfpt}. (b) Vertex part of the self--energy from \e{e:Mep5}: this is
the driving mechanism of the exciton--phonon coupling as explained the text. The vertex function is written in terms of the elemental/irreducible excitonic
propagator $\uu{L}^{el}$ and the statically screened electronic interaction $\uu{W}$ (wiggly line).
}
\label{fig:Mep}
\end{center}
\end{figure}

\lab{ss:Mep}
%=============================================================================================================================
The {\em exact}\cite{VanLeeuwen2004} electron--phonon self--energy is:
\ml{
 \lab{e:Mep1.main}
 M^{e-p}_{ij}\ttw{1}{2}=i \sum_{\mu}G_{lm}\ttw{1}{3} \times \\ 
 \times \(\uu{\uu{\gC}}^{el}\tth{3}{2}{4} \uu{\uu{D}}^\mu\ttw{4}{1}\)\mind{mj}{il}.
}
\e{e:Mep1.main} is written in the reference single particle basis defined in \e{e:T3}, $G$ is the single--particle GF, while $D$ is the {\em dressed} phonon propagator and 
$\gC^{el}$ is the {\em elemental}/irreducible vertex function, given by:
\seq{
 \lab{e:Mep2}
\eqg{
 D^\mu\mind{ij}{kl}\ttw{1}{2}= g^\mu_{ik}\(t_1\) D^\mu\ttw{1}{2} g^\mu_{lj}\(t_2\),\\
 \gC^{el}\mind{mj}{np}\(t_3,t_2;t_4\)=-\frac{\gd G_{mj}\ttw{3}{2}}{\gd U^{tot}_{np} \(t_4\)}.
}}
In \elab{e:Mep2}{a} $g^\mu\(\rr,t\)$  is the {\em dressed} and time--dependent electron--phonon interaction along the phonon normal mode direction $\mu$.
Here we calculate it with state--of--the--art first--principles Density Functional Perturbation
Theory (DFPT), where the matrix elements are given by\cite{Giustino2017}
\ml{
\lab{e:g_dfpt}
g^\mu_{ij}\(t\)\approx g^\mu_{ij}=\frac{1}{\sqrt{2\Omega_{\mu}}}\int \dint \rr_1 \rr_2 \oo{\phi_i\(\rr_1\)}\times \\ 
  \times \varepsilon^{-1}_{Hxc}(\rr_1,\rr_2)\partial_{\mu}\bigr|_{eq} V_{e-ion}(\rr_2)\phi_{j}(\rr_1).
}
In this expression, $\Omega_{\mu q}$ are the ``adiabatic'' phonon frequencies that may be computed in DFPT and are already renormalised by the static $V_H
+V_{xc}$ interaction, $\{\phi_{i}\(\rr\)\}$ are the Kohn--Sham eigenfunctions of the DFT electronic problem, and 
finally $\varepsilon^{-1}_{Hxc}$ is the static dielectric function again describing the screening of lattice vibrations by the interacting electronic system.
In \fig{fig:Mep} the dressed $\uu{g}$ is represented by the filled box ($\blacksquare$).

The GF appearing in \e{e:Mep2} is the exact one, solution of \e{e:T3} with self--energy including $M^{e-p}$. This is a self--consistent problem
requiring a resummation to all orders the e--e and e--p interactions. Here we are interested, however, in the lowest order exciton--phonon scattering. This is
consistent with \ai\, approaches based on the Debye--Waller plus Fan--Migdal approximations\cite{Giustino2017}.

In the present context we can linearize \elab{e:Mep2}{b} by approximating the $\uu{G}$ appearing on the r.h.s. with the electronic one, solution of the Dyson
equation within the SEX approximation. This is a crucial approximation as it corresponds exactly to the excitonic vertex:
\eq{
 \lab{e:Mep3}
 \gC^{el}\mind{ij}{kl}\ttw{3}{4}\equiv \frac{\gd M^{SEX}_{ij}\(t_3\)}{\gd U^{tot}_{kl} \(t_4\)}.
}
A key step now is to rewrite $\gC^{el}$ in terms of the excitonic propagator. Indeed by using Eqs. \eqref{e:S1} and \eqref{e:G1} we get
\eq{
 \lab{e:Mep3.main}
  \gC^{el}\mind{ij}{kl}\ttw{1}{2}=\gd_{ik}\gd_{jl}+i W\mind{ik}{mn} L^{el}\mind{mn}{kl}\ttw{1}{2}.
}
Thanks to \e{e:Mep3.main} the e--p mass operator can be finally rewritten in terms of $L^{el}$:
\ml{
 \lab{e:Mep5}
M^{e-p}_{ij}\ttw{1}{2}=i   \sum_{\mu} G_{lm}\ttw{1}{3}\left[ \left( \uu{\uu{\gd}}\gd\(t_2-t_4\) +       \right.\right. \\
 \left.\left. + \uu{\uu{W}}\,\uu{\uu{L}}^{el}\ttw{2}{4}    \right)            \uu{\uu{D}}^\mu\ttw{4}{1}\right]\mind{mj}{il},
}
\e{e:Mep1.main} and \e{e:Mep3.main} are essential ingredients of our theory and require further discussion. 
The diagrammatic form of \e{e:Mep1.main} is shown in \fig{fig:Mep}.
It is already clear that the appearance of
$L^{el}$, via \e{e:Mep3.main}, in $M^{e-p}$, will play a crucial role in the following of the theory. 
The question is, then, if it were possible to define
$M^{e-p}$ in terms of $L^{opt}$. The answer is no and the reason is that the $\h{V}_{e-ion}$ potential appearing in \e{e:h1} is bare, by definition.
The phonon propagator calculated within  state--of--the--art, first--principles DFPT is defined in terms of the {\em screened} variation of the ionic potential, as seen before.
The inverse dielectric function appearing in \e{e:g_dfpt} is, in DFPT, approximated with the static DFT inverse response functions. As discussed in
Ref. \onlinecite{Marini2015} the dressing of $V_{e-ion}$ absorbs the exchange scatterings from the vertex function and this is the reason why in \e{e:Mep1.main} the elemental vertex appears.

\subsection{The exciton--phonon kernel}\lab{ss:EP}
%=============================================================================================================================
Equation \eqref{e:Mep3.main} connects the e--p mass operator to the \textit{elemental} excitonic GF. 
This is a key property that permits to calculate correctly the exciton--phonon
interaction. 
The next step now is to link $M^{e-p}$ to the \OBSE and its associated response function. 
To this end we go back to the two--times electron--hole GF,
\eq{
 \lab{e:Lopt}
 L^{opt}\mind{ij}{kl}\ttw{1}{2}\equiv \frac{\gd G_{ij}\(t_1,t_1^+\)}{\gd U^{ext}_{kl} \(t_2\)},
}
and to its corresponding response function $\uu{\uu{\chi}}^{opt}\ttw{1}{2}\equiv -i \uu{\uu{L}}^{opt}\(t_1,t_2\)$.
The equation of motion for $L^{opt}$ can be derived with the Schwinger approach, i.e., by using the functional
derivatives to manipulate \e{e:Lopt}, exactly as we did for the time--dependent generalized BSE in Sec.\ref{s:GBSE}.\cite{Attaccalite2011}

This procedure, along with the help of \e{e:Mep5} and \elab{e:G6}{a}, leads to the e--p contribution to the generalized BSE kernel. Indeed we have that
\seq{
\lab{e:EP1}
\ml{
 \tens{\Xi}\tfo{4}{5}{6}{7}=\tens{\Xi}^{e-e}\tfo{4}{5}{6}{7}\\+\tens{\Xi}^{e-p}\tfo{4}{5}{6}{7},
}
and,
\eq{
\tens{\Xi}^{e-e}\tfo{4}{5}{6}{7}=i\gd\ttw{1}{2}\gd\ttw{4}{3^+}\tens{W},
}
}
the last line being the same as that of \e{e:S4} already obtained in the previous Section.
$\tens{\Xi}^{e-p}$ can then be also obtained by functionally deriving \e{e:Mep5} with respect to $\mat{G}$. We see that $\mat{M}^{e-p}$ depends on $G$ via $G_{lm}$,
$\mat{W}$, $\tens{L}^{el}$ and $\tens{D}$. 

We now adopt the usual approximation used to derive the \OBSE\, where $\frac{\gd\tens{W}}{\gd \mat{G}}$  is neglected. In addition we assume that the phonon
propagator is calculated with DFPT, which implies that also $\frac{\gd D}{\gd \mat{G}}=0$. Those approximations do not alter the main finding of this section.\footnote{Indeed, even by performing the two neglected functional derivatives it is possible to show that they lead to the renormalization of the electron--phonon and electron--electron 
interaction. In diagrammatic language this means that they produce  diagrammatic geometries that can be reduced with respect to a $W$ and/or $D$ internal
propagator.}

\begin{figure}%[H]
\begin{center}
\parbox[c]{8cm}{
\begin{center}
 (a)\\
\epsfig{figure=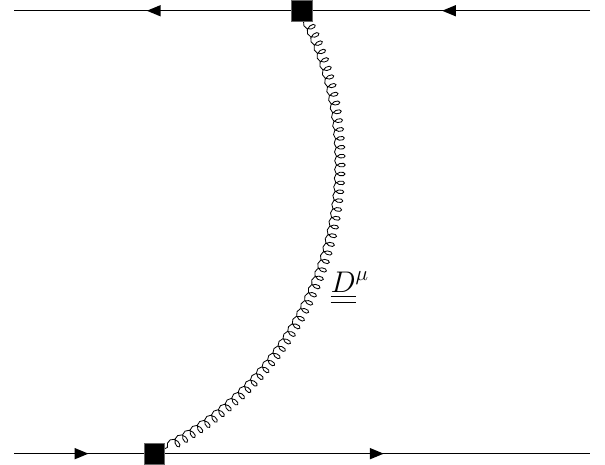,width=0.6\columnwidth}
\end{center}
}
\parbox[c]{8cm}{
\begin{center}
(b)\\
\epsfig{figure=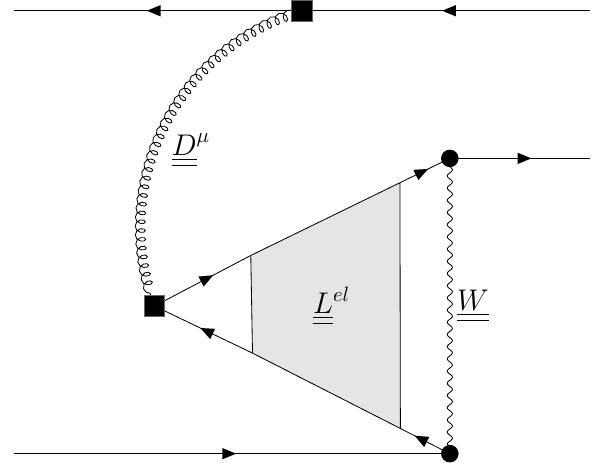,width=0.6\columnwidth}
\end{center}
}
\caption{The two terms contributing to $\left.\Xi^{e-p}\right|_{G}$, \e{e:EP3}. 
Note that diagram (b) represents a renormalization of the electron--phonon vertex.
}\label{fig:XiG}
\end{center}
\end{figure}

We are left, therefore, with two terms
\ml{
\lab{e:EP2}
\tens{\Xi}^{e-p}\tfo{4}{5}{6}{7}=\left.\tens{\Xi}^{e-p}\tfo{4}{5}{6}{7}\right|_{G}\\
+\left.\tens{\Xi}^{e-p}\tfo{4}{5}{6}{7}\right|_{L}.
}
The first term of \e{e:EP2} is easy to evaluate 
\ml{
\lab{e:EP3}
\Xi^{e-p}\left. \mind{ij}{kl}\tfo{1}{2}{5}{6}\right|_{G}=\\
i \sum_{\mu} \[ \(\uu{\uu{\gd}}\gd\(t_2-t_4\)+\uu{\uu{W}}\,\uu{\uu{L}}^{el}\ttw{2}{4}\)\uu{\uu{D}}^\mu\ttw{4}{1}\]\mind{lj}{ik},
}
and is graphically represented in \fig{fig:XiG}. From the diagrammatic representation we see that $\left.\Xi^{e-p}\right|_{G}$, physically, corresponds to a
free electron--hole pair exchanging a phonon lines. It is, in practice, a phonon--mediated electron--hole scattering term.
This scattering occurs at the level of free electron--hole pairs via a DFPT e--p interaction potential, \fig{fig:XiG}(a), and
with a renormalized interaction caused by the exciton--level electron--hole scatterings embodied in $\uu{\uu{L}}^{el}$, \fig{fig:XiG}(b).

\begin{figure}%[H]
\begin{center}
\epsfig{figure=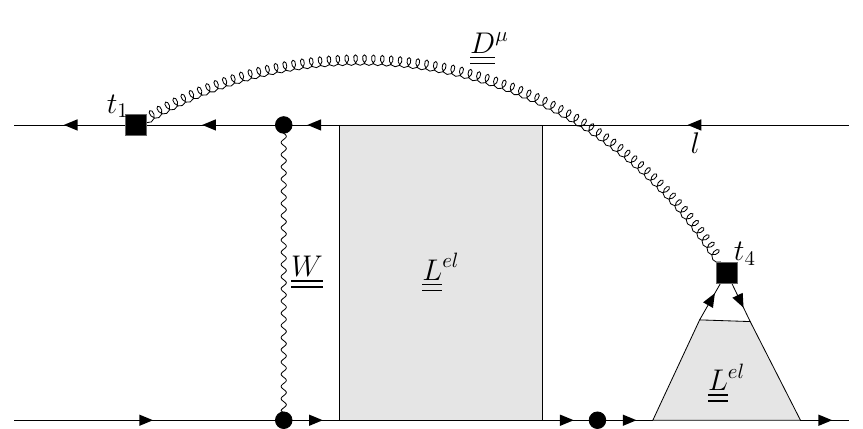,width=\columnwidth}\\
\caption{Graphical representation of the most important diagram contributing to the exciton--phonon kernel. This arises from the change in the \unde GF and it
splits $L^{el}$ in two terms. One re-creates the internal exciton (four point function) while the other dresses the e--p interaction.}\label{fig:XiL}
\end{center}
\end{figure}

The second term of \e{e:EP3} to be calculated is $\left.\Xi^{e-p}\right|_{L}$. This corresponds to the functional derivative of $\uu{\uu{L}}^{el}$ with respect to $G$.
In order to proceed we observe that, by definition,
\eq{
\lab{e:EP4}
\frac{\gd \uu{\uu{L}}^{el}\ttw{2}{4}}{\gd \uu{G}\ttw{5}{6}}=\uu{\uu{L}}^{el}\ttw{2}{7}\frac{\gd\[ \uu{\uu{L}}^{el}\ttw{7}{8}\]^{-1}}{\gd \uu{G}\ttw{5}{6}}
\uu{\uu{L}}^{el}\ttw{8}{4}.
}
In order to calculate $\frac{\gd\[L\]^{-1}}{\gd G}$ we notice that from the equation of motion for $L^{el}$, \e{e:S6}, it follows that
\eq{
\lab{e:EP5}
\[\uu{\uu{L}}^{el}\ttw{7}{8}\]^{-1}=\(\uu{G}\ttw{8}{7}\uu{\uu{\gd}}\uu{G}\ttw{7}{8}\)^{-1}-i\uu{\uu{W}}.
}
Hence, it also follows that
\begin{widetext}
\eq{
\lab{e:EP6}
\frac{\gd \uu{\uu{L}}^{el}\ttw{2}{4}}{\gd \uu{G}\ttw{5}{6}}=\uu{\uu{L}}^{el}\ttw{2}{7}
\[ \uu{\uu{L}}^{0}\ttw{7}{6}\]^{-1}\[\uu{\uu{\gd}}\,\uu{G}\ttw{6}{5}+\uu{G}\ttw{6}{5}\uu{\uu{\gd}}\]\[ \uu{\uu{L}}^{0}\ttw{5}{8}\]^{-1}
\uu{\uu{L}}^{el}\ttw{8}{4}.
}
\end{widetext}
Equation \eqref{e:EP6} may appear complicated but it actually encodes in mathematical form a simple diagrammatic procedure. Indeed the convolution of
$\uu{\uu{L}}^{el}$ with $\[ \uu{\uu{L}}^{0}\]^{-1}$  corresponds to the action of removing the last free electron--hole pair
propagator from the series of diagrams that build $\uu{\uu{L}}^{el}$. 
In this way we get the final form of $\left.\Xi^{e-p}\right|_{L}$:
\ml{
 \lab{e:EP7}
 \left.\Xi^{e-p}\mind{ij}{kl}\tfo{1}{2}{5}{6}\right|_{L}=\\
 = i G_{nm}\ttw{1}{2} \[W\mind{mj}{pq}\frac{\gd L^{el}\mind{pq}{rs}\ttw{2}{4}}{\gd G_{kl}\ttw{5}{6}} 
  D^\mu\mind{rs}{in}\ttw{4}{1}\].
}
\e{e:EP7} leads to two contributions. The first is represented in \fig{fig:XiL} while the second corresponds to the dressed electron--phonon scattering at time
$t_4$ to be positioned on the upper right propagator (index $l$ in the figure) instead that on the lower right one.

The diagrams we have derived in this Section correspond to the so--called ``left'' mass--operator~\cite{Strinati1988}, shown in \fig{fig:Mep}. 
Actually, in order to obtain the complete kernel of the
BSE, we need to apply the same procedure to the right (or adjoint) mass--operator. The difference with \e{e:Mep1.main} is that the vertex $\gC^{el}$ appears on the left side
of the diagram, with one leg at time $t_1$.  
The derivation is exactly the same as we have already done, the only change being the reflection of the ``left'' diagrams with respect to the central time. In this way we obtain four contributions to $\left.\Xi^{e-p}\right|_L$.
\begin{figure*}%[hbtp]
  {\centering \includegraphics[width=0.7\textwidth]{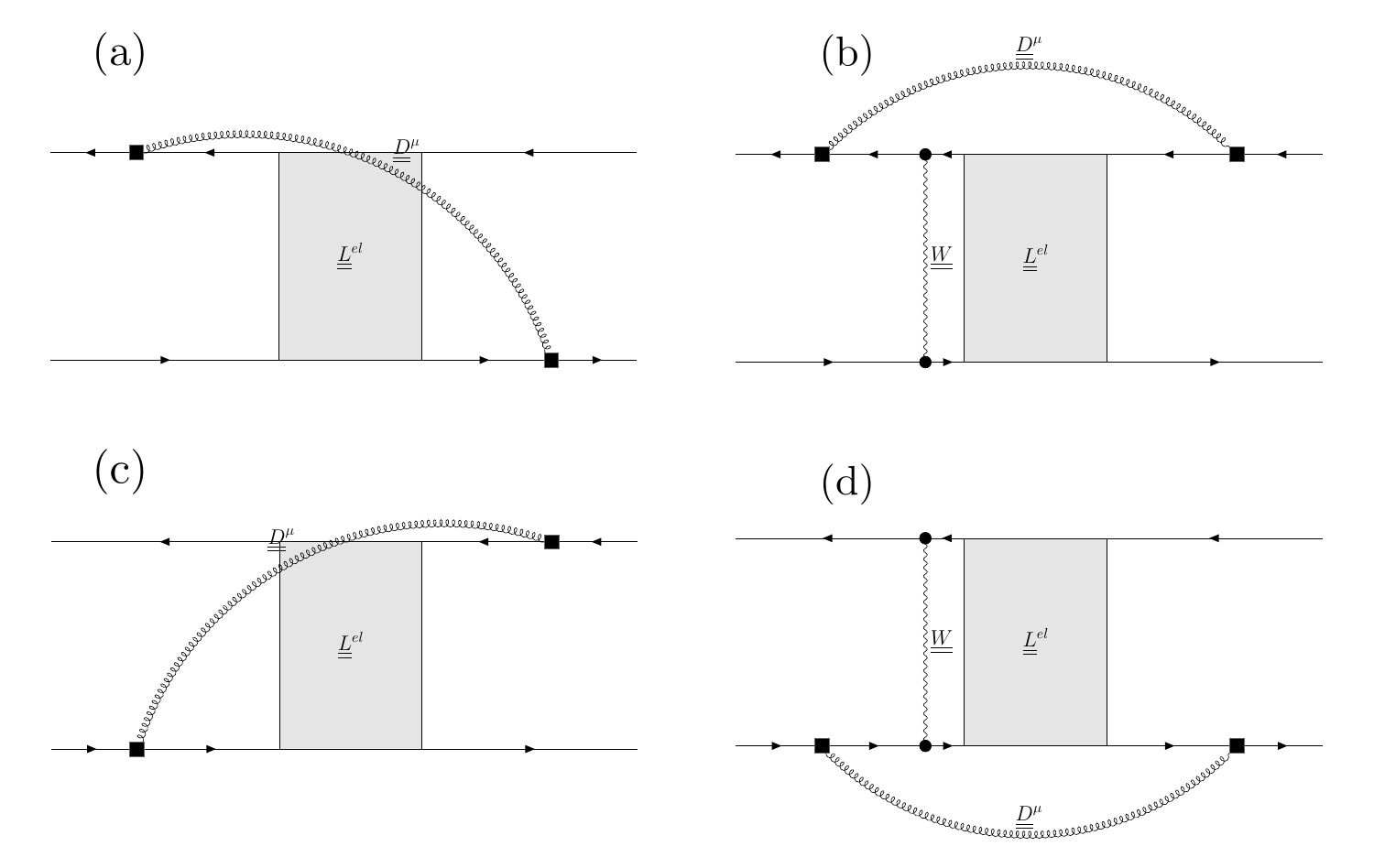}}
  \caption{Final form of the exciton--phonon kernel. This is composed by four diagrams corresponding to the different possible geometries of the phonon
  scattering.}
  \label{fig:Xifinal}
\end{figure*}

An additional crucial approximation that we need to reach the final form of the $\Xi^{e-p}$ is to neglect all internal dressings of the e--p vertexes. The
physical motivation is that the DFPT approach already embodies a correlation contribution due to the DFT exchange--correlation kernel and in order to avoid
double--counting errors the  $\gC^{el}$ renormalization must be neglected. Mathematically this correspond to assume in \fig{fig:XiG} and \fig{fig:XiL} 
$\uu{\uu{\gC}}^{el}\uu{\uu{D}}^\mu\approx\uu{\uu{D}}^\mu$ and to replace the interaction at time $t_4$ in \fig{fig:XiL}  with a plain dressed vertex
($\blacksquare$). At the same time this approximation corresponds to taking only the diagram\,(a) of \fig{fig:XiG}.

Ultimately, if we now sum $\left.\tens{\Xi}^{e-p}\right|_G+\left.\tens{\Xi}^{e-p}\right|_L$ we finally get the final result that is shown diagrammatically in
\fig{fig:Xifinal}. We note that in order to rebuild the full $\uu{\uu{L}}$, the diagrams \figlab{fig:Xifinal}{b} and \figlab{fig:Xifinal}{d} are to be summed with the diagram \figlab{fig:XiG}{a} and its adjoint. 

The final step is to move from time to frequency domain. By using the Feynman diagrams ``cutting'' techniques introduced in
Refs. [\onlinecite{Cudazzo2020b,Marini2003}], it is possible to demonstrate that the Generalized BSE acquires the form
\ml{
 \lab{e:Mep7.main}
 \uu{\uu{L}}^{opt}\(\go\)= 
 \tens{L}^{0}\(\go\)\biggl\{ \tens{1}+\\+i\(\uu{\uu{W}}-\uu{\uu{V}}^{H}\)+\tens{\Xi}^{e-p}\[\uu{\uu{L}}^{el}\]\(\go\)\biggr\}\tens{L}^{opt}\(\go\).
} 
\e{e:Mep7.main} is the main result of this work. The $\tens{\Xi}^{e-p}$ kernel is a functional of the {\em elemental} excitonic propagator. This is a key
property overlooked by previous attempts at deriving a theory of exciton--phonon coupling~\cite{Antonius2022,Chen2020}.

\e{e:Mep7.main} demonstrates what could already be expected from the form of $M^{e-p}$: the {\em optical} exciton--phonon interaction
should be written in terms of \undes. This is not a consequence of the specific diagrammatic form of the mass operator but it is an intrinsic property of bound
electron--hole pairs in a many--body treatment.

\section{Exciton-phonon in practice}\label{s:sim}
%%%%%%%%%%%%%%%%%%%%%%%%%%%%%%%%%%%%%%%%%%%%%%%%%%%%%%%%%%%%%%%%%%%%%%%%%%%%%%%%%%%%%%%%%%%%%%%%%%%%%%%%%%%%%%%%%%%%%%%%%%%%%%
We can now rotate \e{e:Mep7.main} in the \optes basis composed by the states $\ket{\gl}^{opt}$ with energy $E^{opt}_\gl$ to get
\ml{
 \lab{e:Mep7.rotated}
 L^{opt}_{\gl_1 \gl_2}\(\go\)=\left. L^{opt}_{\gl_1}\(\go\)\right|_{HSEX}\[\gd_{\gl_1 \gl_2}+\nl+\Pi_{\gl_1 \gl_3}\(\go\) L^{opt}_{\gl_3 \gl_2}\(\go\)\],
}
with
\eq{
\Pi_{\gl_1 \gl_2}\(\go\)=\sum_{\nu \mu} \frac{\oo{\mc{G}^{\mu}_{\gl^{opt}_1\nu^{el}}}\mc{G}^{\mu}_{\gl^{opt}_2\nu^{el}}}{\go - E^{el}_{\nu}-\Omega_{\mu}-i 0^+}
\lab{e:Pi}
}
being the exciton--phonon self--energy.
In \e{e:Pi} we have defined the {\em optical}--{\em elemental} phonon--mediated scattering potential
\eq{
\lab{e:G}
\mc{G}^{\mu}_{\gl^{opt}\nu^{el}}\equiv \sum_{ijl} \(A_\gl^{ij,opt} \oo{A_\nu^{il,el}}g^\mu_{jl}-A_\gl^{ij,opt}\oo{A_\nu^{lj,el}}g^\mu_{il}\).
}

%\ml{
%\CG_{\lambda\ti{\lambda}^\p q}^{\mu q^\p} = \sum_{cvk}  \[\sum_{v^\p} \(\ti{A}_{\ti{\lambda}^\p q+q^\p}^{ck,v^\p k-q-q^\p}\)^* A_{\lambda q}^{ck,vk-q} g^{\mu q^\p}_{vk,v^\p k-q^\p} \right. \\
% -\left.\sum_{c^\p}\( \ti{A}_{\ti{\lambda}^\p q+q^\p}^{c^\p k+q^\p,vk-q}\)^* A_{\lambda q}^{ck,vk-q} g^{\mu q^\p}_{c^\p k+q^\p,v k} \],
%}
%coupling reducible (BSE\@ HSEX) with irreducible (BSE \@ SEX) excitons via phonons ($\partial V_{e-ion}$ \@ Hxc).

%At this point we remark that with our treatment we have partially recovered the results obtained in Ref. [\onlinecite{Cudazzo2020b,}], except for one major
%differences. First, the present derivation starts from ``scratch'', i.e., from \e{e:h1} including e--p interactions. Second, as a consequence of this, we obtain
%an optical--elemental excitonic scattering (opt--el) instead of the optical--optical (opt--opt) picture so far assumed --- i.e., replacing $\nu^{el}$ with
%$\nu^{opt}$ in \e{e:G}. 
%This is the main result of this work, which goes beyond the usual bosonised exciton model defined in Eqs. \eqref{e:Hbos.1} and \eqref{e:Hbos.2}.  As mentioned
%before, the difference between the opt--el and opt--opt pictures will be negligible in materials where $|V^H|<<|W^{SEX}|$, while it may be sizable in systems
%where this does not hold. 
Eq. \eqref{e:G}, which differs from the commonly employed optical--optical version of the scattering potential, is the central result of this work for computational purposes.
It shows that lattice vibrations can distinguish between total and external fields when coupling to electronic excitations.

\begin{figure}%[hbtp]
  {\centering \includegraphics[width=\columnwidth]{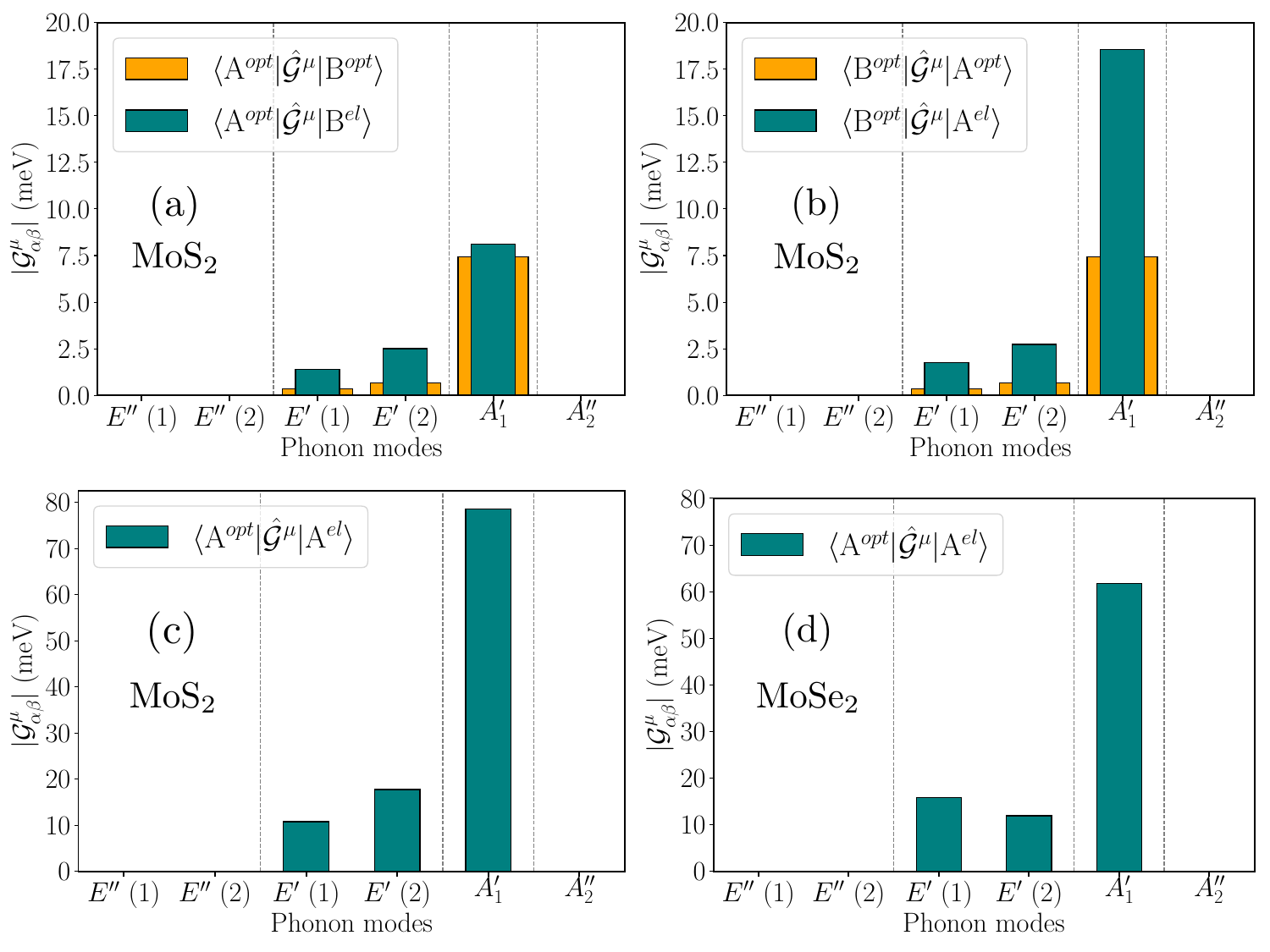}}
\caption{Strengths of the exciton--phonon matrix elements $\CG$ between A and B excitons in \mos and \mose with respect to the optical phonon modes at $q=0$
(the phonon modes are labelled according to their symmetry representation). (a) and (b): comparison between opt--opt (orange) and opt--el (teal) scattering
pictures in \mos. The couplings of exciton A$^{opt}$ with B$^{opt/el}$ (a) and of exciton B$^{opt}$ with A$^{opt/el}$ (b) are shown. (c) and (d): values of the
A$^{opt}$--A$^{el}$ couplings in the case of \mos and \mose, respectively.}\label{f:G}
\end{figure}

In order to complement these equations and remarks with numerical analysis, we consider the excitonic self--energy to be diagonal.
In addition, it is useful to rewrite the self--energy and the coupling matrix element exposing their momentum structure:
\eq{
\lab{e:Pi_q}
\Pi_{\alpha^{opt} q}(\omega) = \frac{1}{N_q}\sum_{\mu\beta^{el} q^\p}\frac{|\CG_{\alpha^{opt} \beta^{el} q}^{\mu q^\p}|^2}{\omega - E_{\beta^{el} q+q^\p}-\Omega_{\mu q^\p}-\im \eta},
} 
and for the coupling we have\cite{Paleari2019t,Chen2020}
\begin{widetext}
\EQ{
\label{e:G_q}
\CG_{\gaopt \gbel q}^{\mu q^\p} = \sum_{cvk}  \[\sum_{v^\p} A_{\gaopt q}^{ck,vk-q} \oo{ A_{\gbel q+q^\p}^{ck,v^\p k-q-q^\p}} g^{\mu q^\p}_{vk,v^\p k-q^\p} -  \sum_{c^\p} A_{\gaopt q}^{ck,vk-q} \oo{A_{\gbel q+q^\p}^{c^\p k+q^\p,vk-q}}  g^{\mu q^\p}_{c^\p k+q^\p,v k} \].
}
\end{widetext}
Here, the $v$ indices correspond to valence (hole) states and the $c$ ones to conduction (electron) states.
We have calculated from first principles the exciton--phonon coupling matrix element, \e{e:G_q}, relative to the A and B excitons of of MoS$_2$ and MoSe$_2$ at
the $\Gamma$ point ($q=0$, $q^\p=0$).  We have done so using both the opt-el ($\CG_{\gaopt \gbel}^{\mu }$) and the opt-opt ($\CG_{\gaopt \beta^{opt}}^{\mu }$)
scatterings in order to compare the two approaches, showing the results in Fig. \ref{f:G} for the A and B excitons for each optical phonon mode $\mu$.
Firstly, we point out that the only active optical modes for these excitons are those with $E^\p$ and $A_1^\p$ symmetry.\cite{Molina2015} Figure \ref{f:G}(a) and (b)
display the values of $|\CG_{A^{opt} B^{\eta}}^{\mu }|$ and $|\CG_{B^{opt} A^{\eta}}^{\mu }|$ for MoS$_2$ with $\eta=opt$ (orange) and $\eta=el$ (teal). This is
not a physically meaningful coupling since it is not possible for exciton A to scatter into exciton B at $\Gamma$ with the aid of a single phonon due to their
energy separation, yet it is useful to compare the opt--opt and opt--el cases.
We see that the couplings are quite similar in terms of active phonon modes, with the opt--el matrix elements having generally a larger value.  Note also that,
as expected, the opt--el matrix at $q^\p=0$ elements are not symmetric in the interchange of the exciton index, unlike in the opt--opt case.  Figures
\ref{f:G}(b) and \ref{f:G}(c) show the couplings between excitons of the same kind, in particular the A states of MoS$_2$ and MoSe$_2$, respectively.  As the
coupling with the $A_1^\p$ mode is close to $100$ meV in both cases, and these scatterings are permitted by energy conservation, we may expect that these matrix
elements, which are absent in the opt--opt case, may play a large role in the excitonic linewidths.\footnote{Note that the $A_2^{\prime\prime}$ mode, i.e., the longitudinal optical mode, is identically zero in our $q=0$ calculation because the long--range Coulomb interaction was not added to the electron--phonon matrix elements\cite{Sohier2016}. This mode is therefore not considered in the evaluation of exciton--phonon transition rates.}

Indeed, from Eqs. \eqref{e:Pi_q} and \eqref{e:G_q} it is possible to define the (homogeneous) linewidth of exciton $\alpha q$ due to phonon scattering.  First,
we switch to the finite temperature version of the self--energy, $\Pi_{\alpha^{opt}q}(w ; T)$, where $T$ is the lattice temperature.  In this case Eq.
\eqref{e:Pi_q}, which describes phonon emission at zero temperature, becomes proportional to $1+n_B(\Omega_{\mu q^\p},T)$, where $n_B$ is the Bose--Einstein
distribution for phonons.  Additionally, a second term appears, this time proportional to $n_B(\Omega_{\mu q^\p},T)$ and describing the phonon absorption
process.  Then, the linewidth can be defined as the imaginary part of the self--energy evaluated at the exciton energy, giving: 
\ml{
\lab{e:widths}
\gamma_{\gaopt q}(T) =
\frac{2\pi}{N_q}\sum_{s\mu\gbel q^\p} |\CG_{\gaopt \gbel q}^{\mu q^\p}|^2 \times \\ \times F_{\mu q^\p}^{(s)}(T)\delta(E_{\gaopt q} - E_{\gbel q+q^\p}-s\Omega_{\mu q^\p}),
}
with $s=\pm$ and $F_{\mu q^\p}^{(s)}(T)= (1+s)/2 + n_B(\Omega_{\mu q^\p},T)$.
A striking occurrence arises from this expression: the linewidth of the optical exciton $\alpha^{opt}$ is now determined by all the elemental excitons $\gbel$ it
can scatter to.  This means, given that elemental excitons are generally more tightly bound than optical ones, that the energy conservation condition
$E_{\ga^{opt}}=E_{\gb^{el}}+\Omega_{\mu}$ may be satisfied even for the lowest--lying optically bright exciton at zero temperature, which therefore
counterintuitively acquires a finite linewidth.  In order to numerically test this, we have computed the $q^\p=0$ component of the linewidths, denoted as
$\gamma^0_{\alpha^{opt}}$, for our excitonic states of interest.  

\begin{figure}%[hbtp]
  {\centering \includegraphics[width=\columnwidth]{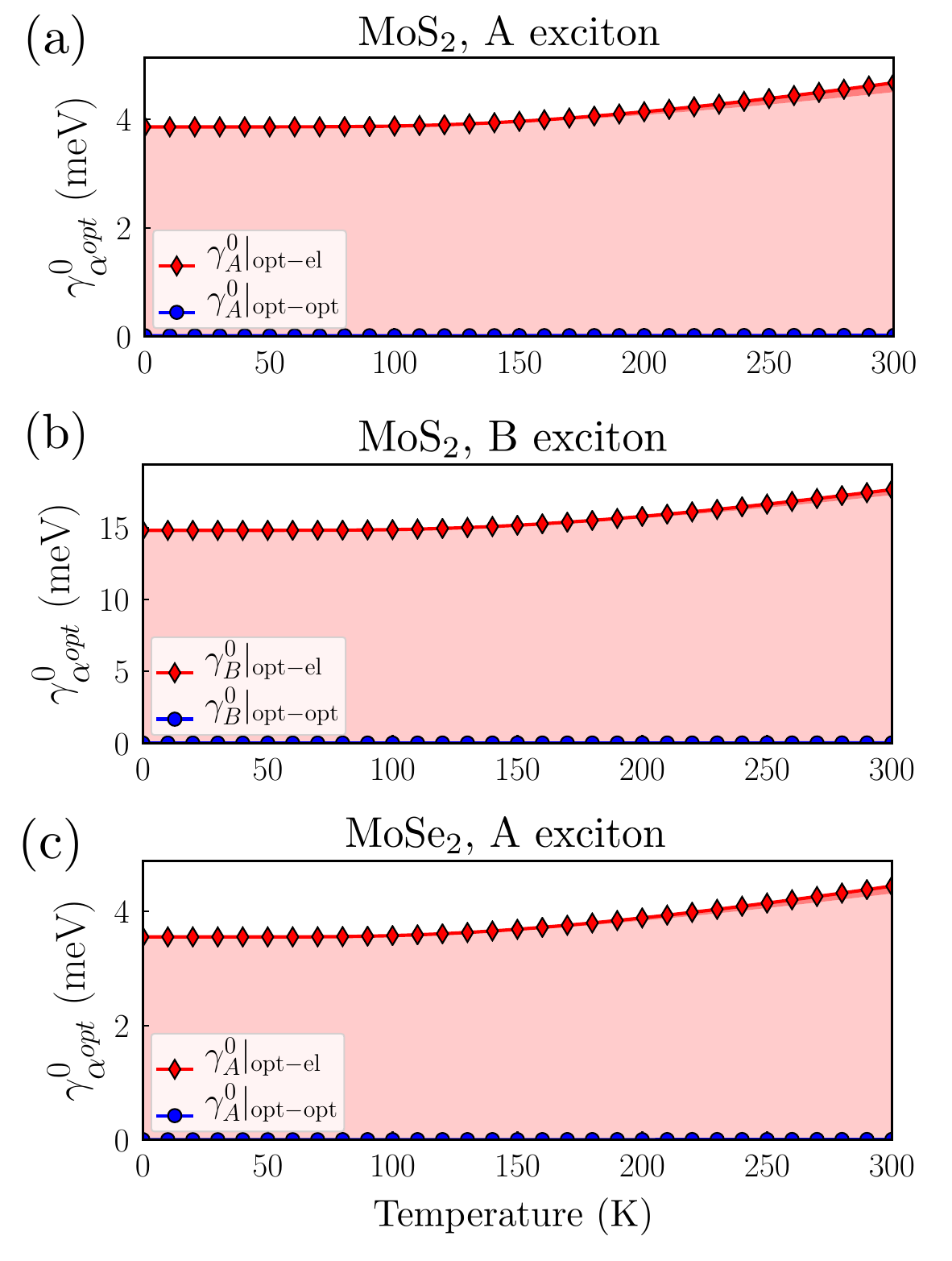}}
\caption{Sum of the exciton-phonon transition rates at vanishing momentum $\gamma^0_{\gaopt}$ as a function of temperature (see text) for select excitonic states in
the optical--elemental scattering picture (red line and diamonds) and in the optical--optical scattering picture (blue line and circles). (a) \mos, A exciton.
(b) \mos, B exciton. (c) A \mose, A exciton. The lightly shaded (darkly shaded, barely visibile on the top) red area represents the phonon emission (phonon
absorption) contribution to $\gamma^0_{\gaopt}$.}\label{fig:excitonic_lifetimes}
\end{figure}

This represents the exciton-phonon transition rates at vanishing momentum and, potentially, the most important contributions to the linewidths due to the large values of the electron-phonon coupling matrix elements for zone-center optical modes in these 2D systems.\cite{Sohier2016,Sio2022,Jin2014}
This quantity is sufficient to assess if the A (and B) excitons have a finite
linewidth or not.
The results for $\gamma^0_{\gaopt}$ are plotted in Fig. \ref{fig:excitonic_lifetimes} as a function of temperature for the A and B excitons of MoS$_2$ ((a) and
(b), respectively) and for the A exciton of MoSe$_2$ (case (c)).  The red diamonds represent the opt--el case, while the blue circles refer to the opt--opt
scatterings (the lightly shaded region corresponds to the contribution of the phonon emission term; the barely visible darkly shaded region is due to the
phonon absorption term).  We see that in the opt--opt case the $q=0$ transition rates are always negligibly small, as expected.  However, in the
case of the opt--el scattering, the same quantity for the A excitons \textit{starts with a finite value} around $4$ meV, even though these states are the
lowest--bound optical excitons in the two systems considered. The B exciton has a much larger value as well.

We note that the opt--el scattering also affects the energy position of the phonon--assisted satellite replicas in absorption or emission spectra\cite{Cudazzo2020b,Paleari2019t,Chen2020} (in addition to their linewidths), since the poles of the self--energy in Eq. \eqref{e:Pi_q} are different than in the opt--opt case --- the difference being $E_{\beta^{opt}} - E_{\beta^{el}}$. For example, the optical absorption satellites relative to exciton $\alpha^{opt}$ (with $q=0$) will appear at energy $E_{\alpha^{opt}} + E_{\beta^{el} q^\prime}-s\Omega_{\mu q^\prime}$ instead of $E_{\alpha^{opt}} + E_{\beta^{opt} q^\prime}-s\Omega_{\mu q^\prime}$.

\section{Discussion}\label{s:discuss}
%%%%%%%%%%%%%%%%%%%%%%%%%%%%%%%%%%%%%%%%%%%%%%%%%%%%%%%%%%%%%%%%%%%%%%%%%%%%%%%%%%%%%%%%%%%%%%%%%%%%%%%%%%%%%%%%%%%%%%%%%%%%%%
In this work we stressed and formally demonstrated the importance of considering ``excitons'' as a product of the measurement process, i.e., as the response of
the electronic system to a specific external experimental field, instead of thinking about them as real particles.  In the case of the exciton--phonon problem,
this leads to the optical--elemental scattering picture rather than the optical--optical picture.  Here we discuss several possible directions of future
investigations.  As for how to compare theory and experiment, at the end of this Section we outline the main underlying challenges.

{\em Excitons as real particles}.  It may be possible, under specific experimental conditions (such as low excitation density, equilibrium--like steady state of
exciton generation and recombination), to reduce the physics of a photo--excited system to simple objects.  However it does not seem possible, at the same time,
to write a theory of exciton dynamics in terms of excitonic boson states.  In general, the difference between the optical--elemental and optical--optical
pictures may be small in materials where $|V^H|<<|W^{SEX}|$, while it may be sizable in systems where this does not hold.  According to our calculations, the
ratios of $|V^H|$ to $|W^{SEX}|$ for the excitons analysed in this paper are: $2.4\%$ (A exciton, monolayer \mos),  $4.0\%$ (B exciton, monolayer \mos), and
$2.8\%$ (A exciton, monolayer \mose). This is sufficient to cause level crossings between bright and dark states and sensibly alter the exciton dispersion
landscape at low finite momenta (see \fig{fig:EL_vs_OPT}).

{\em Photo--luminescence}.  In general it is well known that the frequency of the emitted light may be different from the one corresponding to the peak
absorption (one striking case is the upconversion effect\cite{Han2018}).  Since photo--luminescence is a recombination process involving real excitations of the
system, it is possible that this is purely governed by elemental states, leading to different energies and linewidths.  In other words, our theory suggests that
the emission linewidths and their time--domain counterparts, ``exciton'' relaxation times, should be different from absorption linewidths and
\textit{decoherence} times (only the latter linewidths having a finite value at $T=0$).  

{\em Ultra--fast physics and excitonic dynamics}.  What dictates the
dynamics following a photo--excitation?  Our hypothesis is that the initial excitation does not project the system in a specific and well--defined (``optical'')
excited state but, rather, to a packet of (``elemental'') excited states, e.g., the energetically accessible section of the distributions in
\fig{fig:EL_vs_OPT}(c--d).  This means that the dynamics that follows depends, unavoidably, from the excitation itself. In addition, and more importantly, the
physics of this evolution cannot be described entirely in terms of \optes.

{\em Comparison between theory and experiment.} Unfortunately, the comparison of ab initio, parameter--less theory and experiment is very difficult in these
kinds of systems.  On the theory side, TMD exciton linewidths and phonon--assisted optical spectra carry unknown uncertainties because they depend on
the relative positions of the many valleys appearing in the exciton dispersions of these materials.  These valleys in turn depend enormously on tiny details of
the underlying electronic band structures, and are affected both by pseudopotential details (DFT level) and quality of the $k$--dependent quasiparticle
approximation employed (many--body level). The severity of these problems increases with the atomic number of the elements forming the TMD compound
(particularly if Se, Te, or W are present).  From the experimental point of view, it is very difficult to disentangle the phonon contribution to the exciton
linewidths from other effects.  In particular, beyond the phonon contributions, linewidths are also compounded by the probability of radiative recombination,
substrate and encapsulation dependence and inhomogeneities due to disorder, defects and strain.  In order to unambiguously check our predictions, we would need
an experimental setup able to measure either exciton absorption linewidths (in the frequency domain) or exciton decoherence times (in the time domain) on a very
clean sample and with an accuracy around $1$ meV / $5$ ps while reliably discarding the radiative recombination contributions.  In recent years, attempts have
been made theoretically and experimentally to quantify the latter
contribution\cite{Palummo2015,Selig2016,Wang2016b,Brem2018,Henriques2021,Roux2021,Cassabois2022}.  In addition, multidimensional optical spectroscopy (MDOS)
experiments\cite{Cundiff2013,Moody2017} have allowed for the extraction of the homogeneous part of the linewidths from the inhomogeneous one, leading in many
cases to a reduction of about one order of magnitude in observed linewdiths between MDOS and photoluminescence
experiments\cite{Moody2015,Dey2016,Guo2019,Martin2020}.  In general, MDOS experiments measure values below $10$ meV for the A exciton linewidths at low
temperatures.  Using these techniques, experimentalists are now able to probe the roles of electron--hole exchange and phonon interactions in the exciton
dynamics in real time\cite{Guo2019,Li2021}.  Despite all the current advances, unresolved differing theoretical and experimental estimates for the A exciton
linewidths of MoS$_2$ and MoSe$_2$ are still present in the cited literature --- with no estimates to our knowledge for the B exciton of MoS$_2$.  For example,
in the case of the A exciton of MoSe$_2$, Ref. [\onlinecite{Martin2020}] finds a broadening of less than $1$ meV --- consistent with what the optical--optical
scattering would predict, while Ref. [\onlinecite{Dey2016}] finds a homogeneous value of around $5$ meV for both MoS$_2$ and MoSe$_2$, proposing intrinsic
electron--phonon interactions as the limiting factor: the latter interpretation is quite consistent with our findings. Both are MDOS experiments.

\section{Conclusion}\label{s:conc}
%%%%%%%%%%%%%%%%%%%%%%%%%%%%%%%%%%%%%%%%%%%%%%%%%%%%%%%%%%%%%%%%%%%%%%%%%%%%%%%%%%%%%%%%%%%%%%%%%%%%%%%%%%%%%%%%%%%%%%%%%%%%%%
We have developed a theory of exciton--phonon coupling starting from the many--body interacting electronic Hamiltonian, in the presence of both an external
electro--magnetic field and electron--phonon interaction. In deriving the theory, we have used only those approximations and assumptions already underlying
first--principles treatments of optical excitations and lattice vibrations, without adding additional ones, such as bosonized excitons, which are often employed
in the literature.
Our main finding is that exciton--phonon interaction fundamentally distinguishes between the responses of the electron system to the external and total fields,
coupling the excitations that describe the first (optical, reducible) and the second (elemental, irreducible).
Using the examples of monolayer \mos\, and \mose\, --- two paradigmatic materials belonging to the highly interesting class of layered semiconductors --- we
have also shown how the exciton--phonon matrix elements and linewidths can be qualitatively different with respect to the case when only a single exciton
``type'' is considered.
Therefore, we believe that our work may be valuable for the interpretation and calculation of various exciton--phonon related phenomena, namely exciton linewidths
and broadening, phonon--assisted absorption, emission and reflectivity measurements in the presence of excitons, and the complex nonequilibrium problem of
exciton dynamics.
On the computational side, a full implementation of exciton--phonon interactions is ongoing in the Yambo code~\cite{Sangalli_2019,AndreaMarini2009}, with the aim of computing accurate linewidths and
luminescence spectra including finite--momentum integrals in the Brillouin zone. 
On the theoretical side, two very promising avenues of research are represented by the derivation of a consistent theory for incoherent exciton relaxation
(including the coherent--to--incoherent crossover) and the investigation of the dynamical effects of electronic screening on the electron--hole--phonon
composite excitation.

\section{Acknowledgements}
We gratefully thank C. Attaccalite, D. Sangalli and M. Zanfrognini for precious insight on the numerical simulations and testing of the finite--$q$ BSE and of
the electron--phonon and exciton--phonon couplings. We are also indebted to M. Palummo for sharing input files for the GW--BSE calculations of \mose based on
Ref. [\onlinecite{Marsili2021}] and for discussions about pseudopotentials.

We acknowledge the funding received from the European Union projects: MaX {\em Materials design at the eXascale} H2020-EINFRA-2015-1, Grant agreement n.
676598, and H2020-INFRAEDI-2018-2020/H2020-INFRAEDI-2018-1, Grant agreement n. 824143;  {\em Nanoscience Foundries and Fine Analysis - Europe} H2020-INFRAIA-2014-2015, Grant agreement n. 654360.

\appendix

\section{External and induced fields}\label{APP:fields}
In \e{e:h1} we have used as external perturbation the {\em total} macroscopic potential $U^{ext}$ which includes the induced potential. This is a mathematical short way to
introduce \optes directly from the response function defined in \e{e:chi1}.  
In reality the Hamiltonian 
contains just the external, bare potential, $V^{ext}$ whose dressing by the local, induced potential should appear dynamically\cite{Yamada2019,Ambegaokar1960,Ehara1982,Cho1999}.

Mathematically this corresponds to define the total Hartree potential as
\eq{
 \lab{app:Vh_tot}
 V^{H,TOT}\(\rr,t\) = \int \di \rr^\p \rho\(\rr^\p,t\) v\(\rr,\rr^\p\),
}
so that  $U^{ext}\(\rr,t\)=V^{ext}\(\rr,t\)+\int d\rr  V^{H,TOT}\(\rr,t\)$.
It follows that \e{e:Vh} contains just the microscopic part of the Hartree potential
\eq{
 \lab{app:Vh}
 V^{H}\(\rr,t\) = V^{H,TOT}\(\rr,t\)- \int d\rr  V^{H,TOT}\(\rr,t\).
}

The exciton--phonon derivation is totally independent on the definition of the external potential. It is in fact possible, without changing anything in the
theory, to take as $U^{ext}$ just the external, experimental field $V^{ext}$, leaving the macroscoping average of $V^{H,TOT}$ together with the rest of the electron--electron interaction terms in \e{e:h1}. In this case the present theory may be said to describe \textit{plasmon}--phonon scattering.

%Observable neutral excitations are linked to the response of the electronic system to the external field. 
%In the case of \optes, which is our main concern here, they are associated with peaks in the macroscopic absorption spectrum of the materials. 
%This, in turn, can be obtained from the response to an external field which also incorporates the macroscopic part of $V^{H,MAC}$, since this term also acts as
%the source of the macroscopic induced electric field.\cite{Yamada2019,Ambegaokar1960,Ehara1982,Cho1999}
%In this case $V^{H,MAC}$ is removed from $W_{e-e}$ since it already appears in the ``external'' potential.

%However, it is understood that in the case of \optes the following replacements, which do not affect the derivation, are considered: 
%\seq{
% \lab{e:etot}
%\eqg{
% U^{ext}\(\rr,t\) \rightarrow U^{ext}\(\rr,t\)+V^{H,MAC}\(\rr,t\),\\
% V^H\(\rr,t\) \rightarrow V^H\(\rr,t\) -V^{H,MAC}\(\rr,t\). 
%}
%}

\section{Mathematical conventions}\label{APP:conventions}
In considering operations among matrices, vectors and tensors we will use the following convention
\seq{
 \lab{e:T5}
\eqg{
 \[\uu{V}\,\uu{\uu{M}}\]\mind{ij}{kl}=V_{im} M\mind{mj}{kl},\\
 \[\uu{\uu{M}}\,\uu{V}\]\mind{ij}{kl}=M\mind{im}{kl} V_{mj},\\
 \[\uu{\uu{M}}\,\uu{\uu{O}}\]\mind{ij}{kl}=M\mind{ij}{pq}O\mind{pq}{kl}.
}
}
In \e{e:T5} we use the Einstein convention that all repeated indices are summed.  Notice also that the generalised
single--particle indices may be transformed in Bloch--state indices by explicitly by replacing $i$ and $j$ with band ($n,m$) and crystal momentum ($k,k^\prime$)
indices:
\seq{
 \lab{e:T6}
\eqg{
 i\rightarrow nk,\\
 j\rightarrow mk^\prime,\\
 \gl \rightarrow \alpha q,
}
}
where the last index $\gl$ refers to the excitonic basis, with $\alpha$ indicating the exciton ``branch'' and $q=k-k^\prime$ its
momentum. We will use the generalised indices as much as possible in order to lighten the equations, switching to the other ones when needed.

\section{The Bethe--Salpeter Hamiltonian}\label{APP:BS_hamiltonian}
The \OBSE Hamiltonian emerges when  the retarded time ordering in $t_1$ and $t_2$ is considered ($t_1>t_2$) and the equations are
solved in frequency space by by applying the Laplace transform to the time difference $t_1-t_2$.

The noninteracting part of the BSE, $\uu{\uu{L}}^{0}\ttw{1}{2}\equiv \uu{G}\ttw{1}{2} \uu{\uu{\gd}}\, \uu{G}\ttw{2}{1}$, can be Laplace transformed and results in 
\eq{
 \lab{e:S7}
 L^{0}\mind{ij}{kl}\(\go\)=i\gd_{ik}\gd_{jl}\(f_j-f_i\)\[\go+i0^+-\(\gee_i-\gee_j\)\]^{-1},
}
where $f_i$ represents a single--particle occupation factor.
Then, \e{e:S5} can be rewritten as
\eq{
  \uu{\uu{L}}^{opt}\(\go\)=\[\(\uu{\uu{L}}^{0}\(\go\)\)^{-1}-i\(\uu{\uu{W}}-\uu{\uu{V}}^{H}\)\]^{-1}.
 \lab{e:S8}
}
\e{e:S8} is an exact way to rewrite the BSE in the case of interacting neutral excitations (optical excitons, our focus here, or plasmons). 
At this point the introduction of an effective Hamiltonian can be done by observing that \e{e:S8} can be rewritten using \e{e:S7} as
\EQ{
\lab{e:APP_OBSE}
  \[\uu{\uu{L}}^{opt}\(\go\)\]^{-1}\mind{ij}{kl}=-i \(f_j-f_i\)^{-1}\(\gd_{ik}\gd_{jl}\(\go+i0^+\)-\mc{H}^{opt}\mind{ij}{kl}\),
}
with
\EQ{
\lab{e:APP_HOBSE}
  \mc{H}^{opt}\mind{ij}{kl}=\gd_{ik}\gd_{jl}\(\gee_i-\gee_j\)-\(f_j-f_i\)\(W\mind{ij}{kl}-K^{opt}\mind{ij}{kl}\).
}
\section{Computational details}\label{APP:comp_details}
In this Appendix we provide extensive computational details regarding our many--body, first--principles simulations\cite{Onida2002,Martin2016} of monolayers
MoS$_2$ and MoSe$_2$.  The density functional theory\cite{dft} (DFT) simulations of the electronic ground state and the Kohn--Sham eigenvalues were done with
Quantum ESPRESSO\cite{Giannozzi2009,Giannozzi_2017} (QE). This code was also used for the density functional perturbation theory\cite{dfpt,dfptgonze} (DFPT)
calculation of the phonon frequencies and electron--phonon matrix elements.
The many--body simulations, using DFT as a starting point, were performed with the Yambo code\cite{Sangalli19}.  They include the use of the G$_0$W$_0$
approximation\cite{Hedin1999,Hybertsen1986} for the quasiparticle corrections to the Kohn--Sham eiganvalues, as well as the state--of--the--art
BSE\cite{Albrecht1998} simulations of excitonic properties. 
%We stress that the aim of these calculations was to complement and test the theory with physically relevant results, not to provide extremely accurate numerical convergence benchmarks. 

\textit{DFT and DFPT.} We used norm--conserving, fully relativistic pseudopotentials\cite{Hamann2013} (GGA--PBE type) and included spin--orbit interaction at all stages of the
calculations, working with spinorial wave functions\cite{Marsili2021}.  Our 2D hexagonal systems have lattice parameter $a=5.90$ (\mos) and $a=6.15$ (\mose)
bohrs, with about $40$ bohrs of vacuum separating repeated copies of the simulation supercells in the $c$ direction. A $2$D Coulomb cutoff technique was used
both at the DFT/DFPT level\cite{Sohier2016}, in order to correctly compute phonon--related quantities at vanishing momentum, and at the many--body stage.  The
kinetic energy cutoff on the wave functions were $140$ Ry (\mos) and $90$ Ry (\mose), and the ground--state charge density was converged in both cases with a
$12\times 12\times 1$ grid of $k$--points in momentum space.  Unoccupied Kohn--Sham bands, phonon frequencies, phonon eigenvectors and the variations in the
self--consistent DFT potential were then computed on this charge density. We checked that both the Kohn--Sham band structures and the phonon dispersion curves
were in agreement with previous calculations.
Electron--phonon matrix elements were computed on a $39\times 39\times 1$ $k$--grid to match the one used for excitons.
 
\textit{Many--body.}
The quasiparticle corrections for \mos were simulated using a simple scissor operator enforcing a rigid shift of the bands by $1$ eV, which was enough for our
purposes. For \mose we used instead the quasiparticle corrections previously calculated in Ref. [\onlinecite{Marsili2021}] at the G$_0$W$_0$ level, where the
relevant details may be found.
We computed the BSE both with (reducible, optical) and without (irreducible, elemental) the exchange contribution to the excitonic kernel, all in the
Tamm--Dancoff approximation\cite{Dancoff1950} (i.e., the kernel includes only resonant or antiresonant electron--hole transitions; this common approximation
usually works well for gapped semiconductors and is a requirement for the exciton--phonon treatment).  We used a dense grid of $39\times 39\times 1$ $k$--points
for both systems.
The RPA static screening was computed with an energy cutoff of $8$ Ry, using $100$ empty states in both cases.  The energy cutoff for the exchange part of the
kernel was set to $60$ Ry (\mos) and $40$ Ry (\mose) when included in the calculations.  The cutoff on the RPA screened interaction was $8$ Ry for both systems.
The electronic transitions included in the BSE kernel were comprised in both cases of the two top valence and the two bottom conduction states, properly
including the spin--orbit splitting at the $K$ and $K^\p$ points in the BZ.  We checked that our calculated optical absorption spectra are in agreement with
existing calculations\cite{Marsili2021}.  For the finite--$q$ BSE calculations, \figlab{fig:EL_vs_OPT}{e--f}, we also checked that our results are in agreement
with existing literature.\cite{Deilmann2017}

\textit{Exciton--phonon.}
The phonon frequencies, eigenvectors and electron--phonon matrix elements were read from the DFPT--QE calculations and converted to the Yambo format.  Then,
these quantities along with the exciton energies and $k$--space exciton wave functions from the BSE--Yambo calculations were combined using the Yambopy
package\footnote{Yambopy is a python pre--postprocessing tool for QE and Yambo. It is currently under development, but a public version of the code is already
available on the Yambo website.} in order to compute the exciton--phonon coupling matrix elements, \e{e:G_q}, at $q=0$.  We note that the capability to compute
various exciton--phonon related quantities, including integrations over $q$, is currently being developed in the Yambo code.  Concerning the $q^\p=0$ component
of the linewidth, $\gamma^0_{\alpha^{opt}}$ in \e{e:widths}, a numerical broadening factor of $1$ meV was used for the delta function. Sixteen excitonic states
were included in the sum: this is enough to converge the value of $\gamma^0_{\alpha^{opt}}$ within $0.01$ meV for the excitonic states considered (note that
many more states may be necessary to converge the real part of the self--energy\cite{Antonius2022}). Each plot in Figs. \ref{f:G} and
\ref{fig:excitonic_lifetimes} is summed over the two components of the doubly degenerate A and B states.

\FloatBarrier
\bibliography{excphon.bib}

\end{document}